\documentclass[aps,prd,twocolumn,superscriptaddress,nofootinbib,amsmath, amssymb]{revtex4-2}

\usepackage{graphicx}
\usepackage{dsfont}
\usepackage{bm}

\begin{document}
	
	
	\title{Emerging entanglement on network histories
	}
	
	\author{Cecilia Giavoni}
	\email{Cecilia.Giavoni@physik.uni-muenchen.de}
	\affiliation{Arnold Sommerfeld Center for Theoretical Physics, Theresienstra{\ss}e 37, 80333 Munich, Germany}
	
	\author{Stefan Hofmann}
	\email{Stefan.Hofmann@physik.uni-muenchen.de}
	\affiliation{Arnold Sommerfeld Center for Theoretical Physics, Theresienstra{\ss}e 37, 80333 Munich, Germany}
	
	\author{Maximilian Koegler}
	\email{M.Koegler@physik.uni-muenchen.de}
	\affiliation{Arnold Sommerfeld Center for Theoretical Physics, Theresienstra{\ss}e 37, 80333 Munich, Germany}
	
	\begin{abstract}
    We show that quantum fields confined to Lorentzian histories of freely falling networks in Minkowski spacetime probe entanglement properties of vacuum fluctuations that extend unrestricted across spacetime regions. Albeit instantaneous field configurations are localized on one-dimensional edges, angular momentum emerges on these network histories and establish the celebrated area scaling of entanglement entropy.

	
	\end{abstract}
	\maketitle
	
	
\section{Introduction}
\begin{figure}[t]
	\centering
\includegraphics[width=0.5\textwidth]{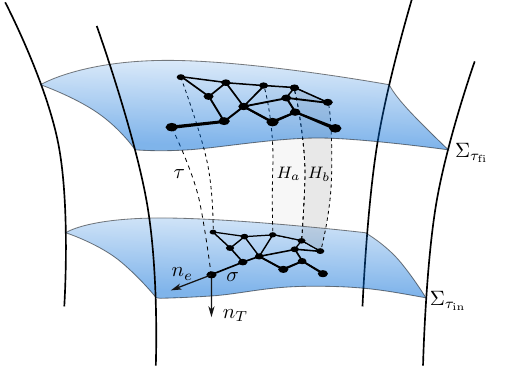}\caption{  \label{spacetime}
        Shown is a portion of spacetime deconstructed
        in accordance with global hyperbolicity. 
        Depicted is an (initial) hypersurface labeled by the
        time $\tau_\mathrm{in}$ assigned 
        to the instantaneous events supported on $\Sigma_{\tau_\mathrm{in}}$
        by freely falling observers.
        On this hypersurface an ideal network  resides. 
        As the network is freely falling to $\Sigma_{\tau_\mathrm{fi}}$
        it traces out a piecewise smooth two-dimensional 
        Lorentzian history to which 
        the dynamical degrees of freedom are confined. As an example, in gray is shown the Lorentzian history traced out by the evolution of two edges of the network.}
\end{figure}

Physical networks consist of communication channels such as coaxial cables, fiber optics or 
telephone lines that connect different vertices of graphlike infrastructures on which 
different hardware units may reside. 
Graphs are an idealization of physical networks, where the vertices are boundary points common to 
multiple communication channels 
represented by one-dimensional edges. In the simplest realization the hardware is replaced 
by certain boundary conditions controlling the transition between edges via the vertices. 
Histories or worldsheets of ideal networks are two-dimensional Lorentzian submanifolds of spacetime
which are only piecewise smooth, that is,
they can be partitioned locally into finitely many smooth
Lorentzian submanifolds so that continuity 
holds across their respective joints (Fig.~\ref{spacetime}).

The idea of confining quantum fields on graphs has been developed in the last two decades and finds its roots in quantum graph theory, according to which differential operators, e.g., Hamiltonians, are confined on metric graphs \cite{berkolaiko2012introduction}. 
On a quantum graph, differential operators act along the edges with appropriate conditions as junction conditions at the vertices. Metric graphs have been the arena to analyze partial differential equations
and appropriate junction conditions, spectral theory of linear operators, quantum chaos and scattering of waves on vertices, for instance \cite{Kottos1, Kottos2, V, fulling2005local, fullingcasimir}.
More recently, the quantum theory of fields confined to graphs was introduced in \cite{Schrader_2009}, 
and a thorough discussion of quantum fields on star graphs has been given in \cite{bellazzini2008quantum, ellazzini_2006}.

In these investigations the metric graph was considered as a fundamental structure. In our new approach, however, fields and differential operators are supported on network histories which are two-dimensional 
piecewise smooth Lorentzian submanifolds embedded in spacetime.
This, in turn, offers the possibility to employ networks as diagnostic devices to probe the embedding spacetime solely by the physics confined to the network histories.
In particular, ideal networks serve as playgrounds to capture physical properties of phenomena supported in 
all of spacetime. 

The advantages of this approach are set by the following remarks: {1.}~Embedding a network in a background induces a conformally flat metric on each of its edges; {2.}~Being confined on the network, the field theory is two-dimensional on each edge history. Hence, the field theory on the network reduces to a sum of simple $(1+1)$-dimensional theories supplied by appropriate boundary conditions at each vertex.

Therefore, solving the $(1+1)$-dimensional theory on a single edge, enables to solve the theory on the whole network. 
This implies that the description of quantum fields is significantly simplified on the network; 
an extremely useful property for investigating phenomena in generic spacetimes, where often a full $(1+3)$-dimensional approach is not accessible.
As a matter of fact, two-dimensional field theories often posses the advantage of being exact solvable, even in some interacting cases or in curved backgrounds \cite{mattis1993many, BELAVIN1984333, Marzlin_2022}. This would provide an exact solution on each edge of the network.

In conclusion, confining (quantum) fields to edges of given (quantum) networks, and thus to two-dimensional 
Lorentzian submanifolds of spacetime, any local observable can be globalized to 
yield an observable on the entire network and its piecewise smooth two-dimensional 
Lorentzian histories which, in turn, allows to investigate the corresponding spacetime phenomena.  
An overview on different possible applications, ranging from curved spacetime frameworks, e.g., black hole physics to gravitational wave detection is presented for the reader in the outlook section.

In this work we employ ideal networks as diagnostic devices to evaluate entanglement 
properties of quantum fluctuations confined to network histories. 
For convenience we choose the entropy of entanglement as a measure for 
the entanglement of vacuum fluctuations. 
It is well-known that 
this measure requires a local finite resolution structure in order to avoid 
a short-distance completion including a fundamental description of gravity \cite{Witten_2018, PhysRevD.34.373}.
Choosing a finite resolution structure amounts to introducing a short-distance scale
corresponding to a minimal separation of the entangling quantum fluctuations. 
In experiments such a scale may be given by the finite width of the border 
across which the entanglement is probed or any other spatial resolution limit
of the hardware infrastructure. Any separation scale 
provided by a characterization of the extrinsic infrastructure
has to be taken into account simply because it grants predictive power, 
and it is not flowing with the renormalization group that takes care 
of the multiscale phenomena created by the dynamical degrees of freedom.

It is essential to distinguish networks from (numerical) lattices even when 
the former are equipped with a regulatory structure. In this work graphs 
are introduced as a physical infrastructure and not as a discretization scheme. 
Furthermore, any physical statement concerning quantum information measures 
or other observables is solely based on studying degrees of freedom 
supported by these infrastructures or their Lorentzian histories 
as two-dimensional submanifolds with boundaries embedded in spacetime. 
In particular, we show how quantum information properties of fields 
in spacetime can be captured by confining these fields on adapted networks
and studying their entanglement properties in a strictly two-dimensional arena. 
In this sense the entanglement properties of fluctuations experiencing 
the full spacetime are an emergent phenomenon of those fields 
confined to the lower dimensional network histories. Alternatively, 
certain networks can conveniently be employed to capture quantum information 
properties that are supported in their embedding geometry. 

\begin{figure}[b]
	\centering
	\includegraphics{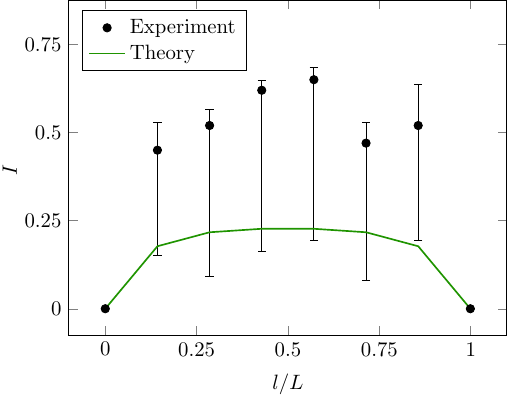}
 \caption{\label{compexp} 
    Shown are the mutual information $I$ of quantum fluctuations 
    confined to a single edge
    in a low-temperature thermal state (data points) and in 
    the ground state at zero temperature (our prediction)
    as a function of the subsystem size $l$ in fractions
    of the edge length $L$.
    }
\end{figure}

Since graphs are idealized networks they can be used to 
describe physical aspects of systems on two-dimensional Lorentzian histories 
provided the idealizations are not interfering with the measurement accuracy
envisaged in dedicated experiments.
A formidable example for this is the recent experimental verification \cite{Tajik_2023} 
of the area law of mutual information by employing effectively two-dimensional
ultracold atom simulators to probe entanglement properties 
predicted in quantum field theory \cite{PhysRevD.34.373} \cite{srednicki1993entropy}.
At zero temperature and for pure states, the mutual information is equal to 
twice the entanglement entropy of either of the subsystems. 
In this experiment the finite resolution structure is determined by the resolution 
of the imaging system which limits access to shorter wavelength modes and 
enforces a short-distance cutoff. A comparison with the data of this experiment 
is shown in Fig.~\ref{compexp}.
It should be noted that the experiment employed a Bose gas at temperatures
between 10 and 100 nano-Kelvin, so the quantum fluctuations
that are accessible in accordance with the regulatory structure are in a 
thermal state resulting in an overpopulation relative to the ground state 
which we assume in our work. 

This article is organized as follows: In Sec.~\ref{secII} we provide a mathematical description of graphs and their histories as idealization of physical networks. Together with the embedding of the network in an arbitrary spacetime, we derive the action and the equations of motion for a scalar field theory on the graph. 
In Sec.~\ref{secIII}, we move to quantum networks, i.e., we analyze quantum properties of a quantum scalar field confined on graphs. As a first investigation, we study the entanglement entropy of vacuum fluctuations confined on minimal graph configurations embedded in Minkowski.
Section~\ref{eeGraph} is finally devoted to the investigation of entanglement entropy of vacuum fluctuations confined on more general networks. The area scaling of the entanglement entropy and its dependence on the shape of the traced out region is investigated. Related discussions and further exciting applications of network histories are finally presented in Sec.~\ref{concl}. Throughout this article, we use the metric signature diag$(-,+,+,+)$ and units where $c=G=\hbar=1$.

\section{Network histories}
\label{secII}

In this work, networks or graphs are meant as collections of objects, called nodes or vertices, and edges
which connect them.
More precisely, networks are ordered triples 
$\mathcal{N}=(N, E, \iota)$, assumed to be irreducible, 
consisting of a finite set $N$ of nodes, a finite set $E$
of edges, and an incident function 
$\iota: E\rightarrow N\times N$ mapping every edge to 
an unordered pair of not necessarily distinct nodes. 
Circumstances permitting, physical networks can be modeled as 
irreducible multigraphs, where the connections are idealized as edges.
For simplicity, physical networks and their idealizations 
will be denoted by the same triple. Throughout this discussion, we explore various perspectives on network configurations. We consider networks with either deformable or rigid edges, those that are in free fall or stationary at a particular location, and networks composed of physical matter or conceptualized as theoretical constructs. Specifically, in this section, we focus on a freely falling network with deformable edges, composed of physical matter, as illustrated in Fig.~\ref{spacetime}.

Let $(M,g)$ be a globally hyperbolic spacetime. 
The finite history of each edge $e$ in $(M,g)$ is a 
two-dimensional compact and connected Lorentzian 
submanifold $(H,h)$ of $(M,g)$, where $h$
is the pullback of the spacetime metric $g$
to $H$. Let $U$ be an open subset of the parameter plane 
$\mathbb{R}^2$ such that horizontal and vertical lines 
intersect $U$ either in intervals or not at all. 
The history or worldsheet of each edge $e$ in $E$ 
is given by a smooth two-parameter map 
$P:U\rightarrow H$, $(\tau,\sigma)\mapsto P(\tau,\sigma)$, 
which is composed of two families of one-parameter 
curves: 
The $\tau$-parameter curve $\sigma=\sigma_0$ of $P$
is $\tau\rightarrow P(\tau,\sigma_0)$, and 
the $\sigma$-parameter curve $\tau=\tau_0$ of $P$ is 
$\sigma \rightarrow P(\tau_0,\sigma)$. 

We can think of the embedding of an edge on $H\subset M$ as a representation 
of this edge on $H$, where its nodes correspond to 
points of $H$ and the edge is a homeomorphic image of the $\sigma$-
parameter curve such that the endpoints of this image coincide 
with the nodes of the edge. 

In this work, networks represent spatial support structures for  
physical degrees  
of freedom in the following sense: The spacetime domains 
of observables 
are assumed to be confined to network histories embedded in spacetime. 

This requires some geometrical preliminaries. Smooth $M$ vector fields $V$ on $H$ 
assign to each point of $H$ a tangent vector to $M$ at this point
such that $V(f)$ is a smooth real-valued function on $H$, provided 
$f$ is a smooth real-valued function on $M$. The set of all such 
vector fields is denoted by $\overline{\mathrm{Vec}}(H)$ and is a module over 
the set of smooth real-valued functions on $H$. The unique 
Levi-Civita connection $D$ of $(M,g)$ induces a connection on $H\subset M$ 
in a natural way. Locally, smooth local extensions
of $V\in \overline{\mathrm{Vec}}(H)$ and $X\in \mathrm{Vec}(H)$ can be constructed, and $D_X V$ can be 
defined using the extended vector fields and then 
restricting the Levi-Civita connection accordingly. 
Since the induced connection  
$D:\mathrm{Vec}(H)\times\overline{\mathrm{Vec}}(H)\rightarrow
\overline{\mathrm{Vec}}(H)$ defined above
is closely related to the Levi-Civita 
connection on $M$, we use the same symbol for both. 

In this section we discuss the covariant theory of classical fields  
confined to a given network infrastructure embedded in 
a globally hyperbolic spacetime. The network metric required for the kinetic operator of these classical fields corresponds to the spacetime geometry 
induced on the network. Hence, the spacetime metric is the only 
classical field that is not confined to the network. 

\subsection{Action and data storage}

Let $\mathcal{N}=(N,E,\iota)$ be a network that is piecewise 
embedded in a globally hyperbolic spacetime $(M,g)$, 
that is each edge in $E$ is embedded in $(M,g)$ 
(subject to boundary conditions). 
Let $n:=\sharp E$ denote the cardinality of $E$.
We equip each edge history $H_a$, $a\in \{1,\cdots,n\}\subset\mathbb{N}$
with a configuration bundle $\mathcal{C}_a$
and, for simplicity, choose real vector bundles of rank one,
$\mathcal{C}_a := (C_a, \pi_a, H_a, \mathbb{R})$. 
Configuration fields on $H_a$ are sections of $\mathcal{C}_a$, 
collectively collected in 
the space 
$\Gamma_\mathrm{c}(H_a,C_a)\subset\Gamma(H_a,C_a)$ 
of compactly supported smooth sections on $H_a$.
Sections of the dual bundle $\pi_a^*: C_a^*\rightarrow H_a$, 
or, equivalently $\mathrm{Hom}(C_a,\mathbb{R}\times H_a)$, 
serve as dual configuration fields. In particular, 
$\Gamma_c(H_a, C_a^*)$ contains linear evaluation forms, 
which are convenient to represent the classical field theory 
in a language close to the one used for the 
corresponding quantum theory. 

Locality is made manifest by a Lagrangian of order $k$, 
$L_a: J^k \mathcal{C}_a^*\rightarrow\Omega_2(H_a)$, 
which is a bundle morphism between the $k$th jet bundle of $\mathcal{C}_a^*$, 
called the Lagrangian phase bundle, 
and the bundle of two-forms over $H_a$.
The action {\it functional} 
$\mathcal{A}_a(\varphi) : \Gamma(H_a,C_a)\rightarrow\mathbb{R}$
is given by
\begin{eqnarray}
    \mathcal{A}_a(\varphi)
    :=
    \int_{H_a} \; 
    s\cdot (L_a\circ j^k \varphi) + \mathcal{B}_a (\varphi)
    \; ,
    \label{action}
\end{eqnarray}
where $\varphi\in \Gamma (H_a,C_a^*)$, $s$ is a 
compactly supported test section in 
$\Gamma_\mathrm{c}(H_a,C_a)$, and $\mathcal{B}_a (\varphi)$
denotes a functional that specifies conditions along the boundary 
$\partial H_a$ of $H_a$. 
Relative to an abstract coordinate system $(x,\varphi)$
in the configuration bundle $\mathcal{C}_a^*$, 
a Lagrangian of order one is represented
by $\mathcal{L}_a(x,\varphi,\mathrm{d}\varphi) \; \mathrm{dvol}_{h_a}$
on the associated coordinate neighborhood, where 
$\mathrm{dvol}_{h_a}$ denotes the natural volume element 
in $(H_a,h_a)$, and the local map $\mathcal{L}_a$ is the usual
Lagrange density.

Consider a vertical vector field $X$ over $\mathcal{C}_a^*$ 
that is required to vanish over the boundary $\partial H_a$, and
denote its flow by $F_t: \mathcal{C}^*\rightarrow \mathcal{C}^*$, 
which is a one-parameter
subgroup of vertical automorphisms of the dual configuration bundle. 
This allows to define a one-parameter family of sections by
$\varphi_t := F_t\circ\varphi$ and consider their actions 
$\mathcal{A}_a(\varphi_t)$. The variation of the action along 
the deformation over $H$ of the action at $\varphi$ is
$(\delta_X \mathcal{A}_a)(\varphi) = \tfrac{\mathrm{d}}{\mathrm{d}t}
\mathcal{A}_a(\varphi_t)|_{t=0}$.
Classical solutions are sections of the bundle $\mathcal{C}_a^*$ at which 
the action is stationary.

The network data is stored as follows: 
Consider a network $\mathcal{N}=(N,E,\iota)$
and let
$H_{\mathcal{N}}=\cup_{a=1}^n H_a$ 
be its connected, piecewise smooth history
relative to a global hyperbolic spacetime $(M,g)$.
Introduce the network action functional 
$\mathcal{A}_{H}$
$:$
$\oplus_{a=1}^n \Gamma(H_a, C_a^*)$
$\rightarrow$
$\oplus_{a=1}^n \Gamma(H_a, C_a^*)$, 
defined by 
$
    \mathcal{A}_H(\varphi^1,\cdots,\varphi^{n})
    :=
    (\mathcal{A}_1(\varphi^1),\cdots, 
    \mathcal{A}_{n}(\varphi^{n}))
$.
For any $F$ in $\oplus_{a=1}^n \Gamma(H_a, C_a^*)$, 
let us write $F = (F^1,\cdots, F^n)$ with $F^a\in \Gamma(H_a, C_a^*)$
and identify $F^a$ with $(0^1,\cdots,F^a,\cdots,0^n)$, where $0^a$ denotes the zero section in $\Gamma(H_a,C_a^*)$. 
In this notation, assuming $A_a(\varphi^a)$ is a functional of degree 
equal to or larger than one, 
$\mathcal{A}_H(\varphi)$ is identified 
with $(\mathcal{A}_H(\varphi^1),\cdots,\mathcal{A}_H(\varphi^n))$
since $\mathcal{A}_H(\varphi^a)
=(0^1,\cdots, \mathcal{A}_a(\varphi^a),\cdots, 0^n)$. 

In order to appreciate networks in the given context, 
we discuss the dynamics of fields 
populating completely disjoint edges and compare 
it to the dynamics on basic building blocks of 
faithful networks. For this, the following concept proves
useful. Let $\langle\cdot,\cdot\rangle_a$ denote the pairing 
$\Gamma_\mathrm{c}(H_a, C_a)\times \Gamma(H_a,C_a^*)$
$\rightarrow$
$\Gamma_\mathrm{c}(H_a,C_a^*)$, where 
$\langle f, F\rangle_a$
is defined to be the integral of $f_a F^a$ 
over the compact support of $f_a$, $\text{supp}(f_a)\subset H_a$
with respect to the canonical measure given by 
the metric $h_a$. 
 
\subsection{Decoupled theory}
Consider a system of $n$ disconnected freely falling edges, 
populated by classical fields. We refer to this as a decoupled theory with action $\sum_{a=1}^n \mathcal{A}_H(\varphi^a)$ with 
$\mathcal{A}_H(\varphi^a)=\mathcal{A}_H^{\; \mathrm{free}}(\varphi^a)$
$+$ $\mathcal{B}_H(\varphi^a)$, where 
$\mathcal{A}_H^{\; \mathrm{free}}(\varphi^a)$ is given by a 
quadratic Lagrangian of order one, describing the free evolution 
on the edge $e_a$, $a\in \{1, \cdots, n\}$, 
and $\mathcal{B}_H(\varphi^a)$ is given by a linear Lagrangian 
of order zero, describing Dirichlet boundary conditions 
on $\partial H_a$ via Lagrange multiplier fields $\lambda_d$, 
$d\in \{1, \cdots, 4\}$. 

The boundary $\partial H_a$ of a finite history $H_a$
can be described as follows: Let $U$ be an open subset
of the parameter plane $\mathbb{R}^2$ and 
$\tau_\mathrm{in},\tau_\mathrm{fi}, \sigma_\mathrm{f}^1, \sigma_\mathrm{f}^2$
be history parameters in $U$ such that
$\partial_\mathrm{f}^1 e_a: [\tau_\mathrm{in},\tau_\mathrm{fi}]$
$\rightarrow H_a$, 
$\tau\mapsto \partial_\mathrm{f}^1 e_a (\tau):= P_a(\tau,\sigma_\mathrm{f}^1)$, 
is the worldline of a free endpoint of edge $e_a$ in the history $H_a$, 
and accordingly for the other free endpoint $\partial_\mathrm{f}^2 e_a$
of $e_a$. Furthermore, 
$\partial_\mathrm{in}T_a: [\sigma_\mathrm{f}^1,\sigma_\mathrm{f}^2]$
$\rightarrow H_a$, 
$\sigma\mapsto \partial_\mathrm{in}T_a(\sigma)$
$:=P_a(\tau_\mathrm{in},\sigma)$ describes the edge $e_a$
at initial parameter time in $H_a$, and, accordingly, 
$\partial_\mathrm{fi}T_a$ gives the same edge at final
parameter time $\tau_\mathrm{fi}$ in $H_a$. 
Now we introduce projections onto the above boundary segments:
For $i\in[2]$, $p\in H_a$, 
$\mathcal{X}_{\partial_\mathrm{f}^i e_a}(p) \cdot \varphi^a(p) (\phi_a)$
$:=\phi_a(p)$ if $p$ lies on the worldline of 
the specified free endpoint of $e_a$ in $H_a$, 
and zero otherwise. 
Furthermore, for $A\in\{\mathrm{in},\mathrm{fi}\}$, $q\in H_a$, 
$\mathcal{X}_{\partial_\mathrm{A}T_a}(q) \cdot \varphi^a(q) (\phi_a)$
$:=\phi_a(q)$ if $q$ is located on $e_a$
at the specified time, and zero otherwise. 
Consider 
$\mathcal{A}_H^{\; \mathrm{free}}(\varphi^a)$
$=$ 
$\langle 
s, -\tfrac{1}{2} h^{-1} (\mathrm{d}\varphi,\mathrm{d}\varphi)
\rangle_a$ with a compact smearing section $s\in \Gamma_\mathrm{c}(H_a, C_a)$, 
where we used the pairing introduced above and 
$\mathcal{B}_H(\varphi^a) =
\langle s, \mathcal{X}_{\partial_\mathrm{f}^1 e} \, \lambda_1 \varphi
\rangle_a
$
$+$
$
\langle s, \mathcal{X}_{\partial_\mathrm{f}^2 e} \, \lambda_2 \varphi
\rangle_a
$
$+$
$\langle s, \mathcal{X}_{\partial_\mathrm{in}T} \, \lambda_3 \varphi
\rangle_a
$
$+$
$\langle s, \mathcal{X}_{\partial_\mathrm{fi}T} \, \lambda_4 \varphi
\rangle_a
$.
These boundary terms enforce Dirichlet boundary conditions for classical fields populating 
$n$ freely falling edges, i.e., 
$\phi_a=0$ on the $2n$ worldlines of the free endpoints, 
and $\phi_a = 0$ for the $2n$ field configurations 
on the edge at the initial and final time. 

The dynamical content requires little more work:
As above, consider a vertical vector field $X_a$ over $C_a^*$. 
Setting $f_a := X_a(\varphi^a)$ and choosing the support functions $s_a$ in 
$\mathcal{A}_H^\mathrm{free}(\varphi^a)$ such that 
$s_a=1$ on the support of $f_a$, we have 
$(X_a \mathcal{A}_H^\mathrm{free}(\varphi^a))(\phi_a) = \langle f, \Box_h \phi\rangle_a + \mathcal{S}_H(\varphi^a)(\phi_a)$.
The surface term is given by 
\begin{eqnarray}
    \mathcal{S}_H(\varphi^a)(\phi_a)
    &=&
    \left\langle
        f, \left[
                \mathcal{X}_{\partial_A T}
            \right]_\mathrm{in}^\mathrm{fi}
            D_{n_T} \phi
    +
            [
                \mathcal{X}_{\partial_\mathrm{f}^i e}
            ]_\mathrm{1}^\mathrm{2} \,
            D_{n_e} \phi
    \right\rangle_a
    \;, \nonumber
\end{eqnarray}
where $n_{T}\perp \partial_\mathrm{in} T$ is chosen to be a past-pointing
vector field normal to the initial edge in $H_a$, 
and 
$n_e$ is a vector field normal to the worldline of $\partial_\mathrm{f}^1 e$
in $H_a$, oriented away from the edge. 
Furthermore, $[\zeta_\nu]^\alpha_\beta := \zeta_\alpha - \zeta_\beta$.   
A straightforward calculation gives for the variation of the boundary action
$(X_a \mathcal{B}_H(\varphi^a))(\phi_a)$
$=$
$\langle f , \mathcal{X}_{\partial_\mathrm{f}^1 e} \lambda_1\rangle_a$
$+$
$\langle f , \mathcal{X}_{\partial_\mathrm{f}^2 e} \lambda_2\rangle_a$
$+$
$\langle f , \mathcal{X}_{\partial_\mathrm{in} T} \lambda_3\rangle_a$
$+$
$\langle f , \mathcal{X}_{\partial_\mathrm{fi} T} \lambda_4\rangle_a$. 

The wave equation for classical fields on a system of freely falling, disjoint 
edges with Dirichlet conditions imposed on the boundaries of the respective 
histories is given by 
\begin{eqnarray}
\label{eom}
    &&\Box_h \phi
    +
    \big[
        \mathcal{X}_{\partial_A T}
        \, D_{n_T} \phi
    \big]_\mathrm{in}^\mathrm{fi}    
    +
    \big[
        \mathcal{X}_{\partial_f^i e}
        D_{n_e} \phi
    \big]_\mathrm{1}^\mathrm{2} 
    \nonumber \\
    &&=
    - \mathcal{X}_{\partial_\mathrm{f}^1 e} \lambda_1
    - \mathcal{X}_{\partial_\mathrm{f}^2 e} \lambda_2
    - \mathcal{X}_{\partial_\mathrm{in} T} \lambda_3
    - \mathcal{X}_{\partial_\mathrm{fi} T} \lambda_4
    \; ,
\end{eqnarray}
where we suppressed the edge label for ease of notation. 
On the interior of each history $H_a$, we find $\Box_h\phi=0$. 
Using this in (\ref{eom}) we can solve for the Lagrange multiplier fields,
\begin{eqnarray}    
    \lambda_1
    &=&
    - \Box_{h} \phi \; \big|_{\partial_\mathrm{f}^1 e}
    - D_{n_e} \phi
    - \big[
        \mathcal{X}_{\partial_A T}
        \, D_{n_T} \phi
      \big]_\mathrm{in}^\mathrm{fi} \; \big|_{\partial_\mathrm{f}^1 e}
      \; , \nonumber \\
    \lambda_2
    &=&
    - \Box_{h} \phi \; \big|_{\partial_\mathrm{f}^2 e}
    + D_{n_e} \phi
    - \big[
        \mathcal{X}_{\partial_A T}
        \, D_{n_T} \phi
      \big]_\mathrm{in}^\mathrm{fi} \; \big|_{\partial_\mathrm{f}^2 e}
      \; , \nonumber \\
    \lambda_3
    &=&
    - \Box_{h} \phi \; \big|_{\partial_\mathrm{in}T}
    + D_{n_T} \phi
    - \big[
        \mathcal{X}_{\partial_\mathrm{f}^i e}
        \, D_{n_T} \phi
      \big]_1^2 \; \big|_{\partial_\mathrm{in} T}
      \; , \nonumber \\
      \lambda_4
    &=&
    - \Box_{h} \phi \; \big|_{\partial_\mathrm{fi}T}
    - D_{n_T} \phi
    - \big[
        \mathcal{X}_{\partial_\mathrm{f}^i e}
        \, D_{n_T} \phi
      \big]_1^2 \; \big|_{\partial_\mathrm{fi} T}
      \; .
\end{eqnarray}
Notice that at the free endpoints, we can rewrite the first two equations as
\begin{eqnarray}    
   D_{n_e} \phi \big|_{\partial_\mathrm{f}^1 e} \
    &=&
   -\lambda_1 + \big[
        \mathcal{X}_{\partial_A T}
        \, D_{n_T} \phi
      \big]_\mathrm{in}^\mathrm{fi} \; \big|_{\partial_\mathrm{f}^1 e}
      \; , \nonumber \\
   D_{n_e} \phi \big|_{\partial_\mathrm{f}^2 e} \
    &=&
   \lambda_2 + \big[
        \mathcal{X}_{\partial_A T}
        \, D_{n_T} \phi
      \big]_\mathrm{in}^\mathrm{fi} \; \big|_{\partial_\mathrm{f}^2 e}
\end{eqnarray}
where the right-hand sides are constants. 
         The above equations guarantee a constant field derivative at the boundaries, thereby enforcing total reflection and consequently preventing any flux leakage from the network. 
\subsection{Coupled theory}
Next, we allow connections of edges into nodes. The minimal network we can consider is a freely falling star graph, that is a network consisting
of $n$ edges each connected to all others 
at a single vertex, and each populated by classical 
fields with action 
$\mathcal{A}_H(\varphi^a)$
$=$
$\mathcal{A}_H^\mathrm{free}(\varphi^a)$
$+$
$\mathcal{B}_H(\varphi^a)$
$+$
$\mathcal{C}_H(\varphi^a)$. A star graph serves as the fundamental building block for more complex networks, hence by introducing the theory for such a minimal junction, we inherently provide a theoretical framework applicable to networks of arbitrary configurations.

Without loss of generality, let the joining vertex $\partial_\mathrm{j} e_a$
be indexed with label j. The boundary action $\mathcal{B}_H(\varphi^a)$
needs to be adapted to this setting, which amounts to  
Dirichlet conditions for the $n$ free endpoints. For finite histories $H_a$, 
the temporal boundary conditions remain the same as in the decoupled theory so instead of having $4n$ Dirichlet conditions, the boundary action now
specifies only $3n$ conditions for the star graph setting. 
The remaining $n$ conditions are provided by the coupling action, 
whereby one is trivial: 
Introduce the two-edge coupling functional by
$\mathcal{C}_{an}$ 
$=$
$[\langle s, \lambda^\mathrm{c} \mathcal{X}_{\partial_\mathrm{j} e} \varphi
\rangle_i]^a_n$. The star-graph 
coupling action is just a sum 
over all edges of these two-edge coupling forms, 
yielding the following coupling conditions 
at the joining node for each adjacent edge $a$: 
\begin{equation}
\mathcal{X}_{\partial_\mathrm{j} e_a} \phi_a
=
\mathcal{X}_{\partial_\mathrm{j} e_n} \phi_n, 
\label{KN}
\end{equation}
that is, the field configurations are continuous 
across the worldline of the joining vertex. 
This specific choice of coupling conditions accounts for the so-called Kirchhoff-Neumann conditions (which reduce to the known Neumann conditions for $n=2$). By imposing continuity of the field configurations, Kirchhoff-Neumann conditions ensure energy conservation at each vertex, which will act neither as a sink nor as a source for the field. In general, different choice of coupling conditions will lead to describing different physical setups. Since in this article we aim to describe physical networks as webs of communication channels, we demand each vertex of the idealized graph to be physically analogous to a node in electrical currents -- for which what enters in has to come out -- condition ensured by the Kirchhoff-Neumann conditions.

The wave equation for classical fields populating 
a star network with Dirichlet boundary and coupling
conditions imposed is given by 
\begin{eqnarray}
\label{eomc}
    &&\Box_{h_a} \phi_a
    + 
    \big[
        \mathcal{X}_{\partial_A T_a} D_{n_{T_a}} \phi_a
    \big]^\mathrm{fi}_\mathrm{in}
    +
    \big[
        \mathcal{X}_{\partial_\mathrm{r} e_a} D_{n_{e_a}} \phi_a
    \big]^\mathrm{f}_\mathrm{j}
    =
    \nonumber \\ 
    &&=
    - \mathcal{X}_{\partial_\mathrm{f}e_a} \lambda_{1a}
    - \mathcal{X}_{\partial_\mathrm{in} T_a} \lambda_{2a}
    - \mathcal{X}_{\partial_\mathrm{fi} T_a} \lambda_{3a}
    +
    \nonumber \\
    && \; \; \; \; - \left(1-\delta_{an}\right) \mathcal{X}_{\partial_\mathrm{j} e_a} 
        \lambda_a^\mathrm{c}
        + \delta_{an} \mathcal{X}_{\partial_\mathrm{j} e_n} 
        \sum_{b\in[n-1]} \lambda_b^\mathrm{c}
    \; .  
\end{eqnarray}
On the interior of $H_a$, Eq.~(\ref{eomc}) reduces to 
$\Box_{h_a} \phi_a = 0$. Using this in (\ref{eomc}), 
we determine the Lagrange multiplier fields associated 
with the Dirichlet boundary segments at the extremities of the star graph,
\begin{eqnarray}
    \lambda_{1a}
    &=&
    -\Box_{h_a} \phi_a \big|_{\partial_\mathrm{f} e_a} 
    - D_{n_{e_a}} \phi_a 
    - 
    \big[
        \mathcal{X}_{\partial_A T_a} D_{n_{T_a}} \phi_a
    \big]^\mathrm{fi}_\mathrm{in} \big|_{\partial_\mathrm{f}e_a}
    \; , \nonumber \\
    \lambda_{2a}
    &=&
    -\Box_{h_a} \phi_a \big|_{\partial_\mathrm{in} T_a} 
    + D_{n_{T_a}} \phi_a 
    - 
    \big[
        \mathcal{X}_{\partial_r e_a} D_{n_{e_a}} \phi_a
    \big]^\mathrm{f}_\mathrm{j} \big|_{\partial_\mathrm{in}T_a}
    \; , \nonumber \\
    \lambda_{3a}
    &=&
    -\Box_{h_a} \phi_a \big|_{\partial_\mathrm{fi} T_a} 
    - D_{n_{T_a}} \phi_a 
    - 
    \big[
        \mathcal{X}_{\partial_r e_a} D_{n_{e_a}} \phi_a
    \big]^\mathrm{f}_\mathrm{j} \big|_{\partial_\mathrm{fi}T_a}
    \; . \nonumber \\
\end{eqnarray}

In addition, for the internal vertex we find the Lagrange multiplier fields associated 
with the coupling of the edges to a star network, 
\begin{eqnarray}
\label{lmfc}
    &&\lambda_a^\mathrm{c}
    =
    -\Box_{h_a} \phi_a \big|_{\partial_\mathrm{j} e_a} 
    + D_{n_{e_a}} \phi_a 
    - 
    \big[
        \mathcal{X}_{\partial_A T_a} D_{n_{T_a}} \phi_a
    \big]^\mathrm{fi}_\mathrm{in} \big|_{\partial_\mathrm{j}e_a}
    \; , \nonumber \\
    &&\sum_{b\in[n-1]} \lambda_b^\mathrm{c}
    =
    \nonumber \\
    &&=
    -\Box_{h_n} \phi_n \big|_{\partial_\mathrm{j} e_n} 
    - D_{n_{e_n}} \phi_n 
    - 
    \big[
        \mathcal{X}_{\partial_A T_n} D_{n_{T_n}} \phi_n
    \big]^\mathrm{fi}_\mathrm{in} \big|_{\partial_\mathrm{j}e_n}
     ,
\end{eqnarray}
for $a\in \{1, \cdots, n-1\}$. Given the network's orientation, 
the second equation in (\ref{lmfc}) is a typical example 
for a collection of conservation laws associated with the internal vertex. 
For instance, power counting arguments allow to extract from (\ref{lmfc})
the following statements:
\begin{eqnarray}
    \Box_{h_n} \phi_n \big|_{\partial_\mathrm{j} e_n} 
    &=&
    \sum_{a\in[n-1]} \Box_{h_a} \phi_a \big|_{\partial_\mathrm{j} e_a} 
    \; , \nonumber \\
    D_{e_n} \phi_n \big|_{\partial_\mathrm{j} e_n} 
    &=&
    -\sum_{a\in[n-1]}
    D_{n_{e_a}} \phi_a \big|_{\partial_\mathrm{j}e_a} 
    \; .
\end{eqnarray}
The last equation can be rewritten as
\begin{equation}
    \sum_{a\in[n]}
    D_{n_{e_a}} \phi_a \big|_{\partial_\mathrm{j}e_a} =0
    \; .
    \label{KNder}
\end{equation}
Together with the smoothness condition for the total field $\Phi := \phi_a \mathbf{e}_a$
across the worldline of the node, this equation ensures that fluxes can propagate across the vertex and that the node does not act as a source or a sink, thereby ensuring energy conservation. As already mentioned, these junction conditions are commonly referred to as the Kirchhoff-Neumann conditions.

The junction conditions Eq.~\eqref{KN} and Eq.~\eqref{KNder} at the vertex j of the graph can be jointly expressed for $\Phi$ as
\begin{equation}
    A \Phi(\mathrm j) + B \Phi^\prime(\mathrm j) = 0\, ,
    \label{bdyCondGen}
\end{equation}
where $A$ and $B$ are complex $n \times n$ matrices and $\Phi$ together with $\Phi^\prime$ are vectors including the field and its derivative for all edges adjacent to the vertex j,
\begin{equation}
    \Phi(\mathrm j) := \begin{pmatrix} \mathcal{X}_{\partial_\mathrm{j} e_1}  \phi_1  \\ \mathcal{X}_{\partial_\mathrm{j} e_2} \phi_2  \\ \vdots  \end{pmatrix}, 
    \quad \Phi^\prime(\mathrm j) := \begin{pmatrix} D_{n_{e_1}} \phi_1 \big|_{\partial_\mathrm{j}e_1} \\ D_{n_{e_2}} \phi_2 \big|_{\partial_\mathrm{j}e_2} \\ \vdots  \end{pmatrix} \, .
\end{equation}
If the $n \times 2n$ composite matrix $(A, B)$ has rank $n$ and $A B^\dagger$ is self-adjoint, then the Laplace operator on the free falling coupled edges is also self-adjoint \cite{V, Schrader_2009}.

As shown above, by requiring the most simple coupling \eqref{KN} for the fields at the vertex,  Eq.~\eqref{KNder} and therefore Kirchhoff-Neumann coupling conditions naturally arise at the vertex with,
\begin{equation}
    A = \begin{pmatrix} 1 & -1 & 0 & \dots & 0& 0 \\ 
     0 & 1 & -1 & \dots & 0 & 0\\
     \vdots & \vdots & \vdots & \ddots & \vdots & \vdots\\
     0 & 0 & 0 & \dots & 1 & -1\\
     0 & 0 & 0 & \dots & 0 & 0\end{pmatrix}\!\!, 
     B = \begin{pmatrix} 0 & 0 & 0 & 0 & 0& 0 \\ 
     0 & 0 & 0 & 0 & 0 & 0\\
     \vdots & \vdots & \vdots & \ddots & \vdots & \vdots\\
     0 & 0 & 0 & \dots & 0 & 0\\
     1 & 1 & 1 & \dots & 1 & 1\end{pmatrix} \! .
     \label{ABkn}
\end{equation}
The above mentioned conditions for $(A,B)$ and $AB^\dagger$ are naturally satisfied such that a self-adjoint Laplace operator is ensured on the graph.

This is crucial since then the completeness relation of the Laplace operator eigenfunctions can be used to construct quantum field and conjugated momentum operators on the star graph that satisfy the equal time commutation relation \cite{V, Schrader_2009}. 
A general network can be manufactured by coupling edges of arbitrary star graphs into new vertices. Smoothness conditions at each vertex, yield again Kirchhoff-Neumann conditions and hence a self-adjoint Laplace operator on the whole network.
This sets the stage for the next section in which we apply standard quantization techniques to define a quantum field confined to the network in order to investigate its entanglement entropy. 

As has been shown in \cite{V, Schrader_2009}, another remarkable consequence of a field theory on the network with a self-adjoint Laplace operator is the unitarity of the scattering matrix, which describes how the field $\Phi$ reflects off and transmits across each vertex. Since the scattering matrix of the whole network can be factorized in terms of the scattering matrix of each star graph, the total scattering matrix is also unitary. 
In terms of a quantum field theory on a network, a unitary scattering matrix ensures the conservation of the probability current across each vertex. In this way, the boundary conditions~\eqref{bdyCondGen} manifest as a quantum version of Kirchoff's law.

\label{sec2}

\section{Entanglement on network histories}
\label{secIII}

\begin{figure}
\centering
\includegraphics[]{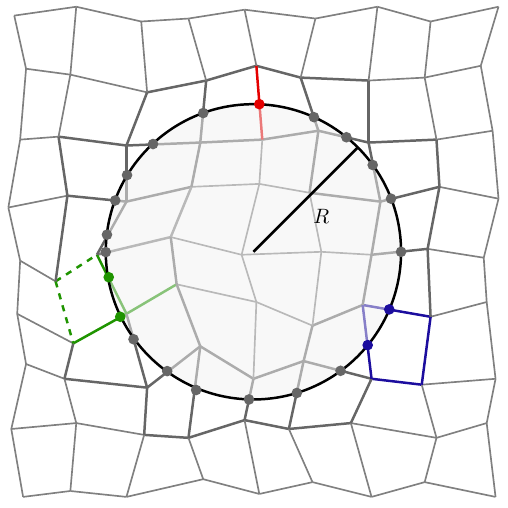}
\caption{\label{meshGraph} 
    Spatial section of a three-dimensional 
    mesh graph embedded in Minkowski spacetime. 
    The disk of radius $R$ is traced out and the crossing points of this circle 
    and the graph are indicated with dots.
    The single edge shown in red, the simple loop in blue, 
    and star graph with three edges depicted in green
    represent basic building blocks of this graph. Moreover, 
    a subgraph consisting of this star graph and two 
    additional edges (dashed) is shown as an example for 
    a loop with a single edge attached.}
\end{figure}   
An example of a general network can be thought of as the mesh graph depicted in Fig.~\ref{meshGraph}.
Evolving edges trace out two-dimensional Lorentzian histories to which 
we assign actions \eqref{action} including boundary specifications 
like the Kirchhoff-Neumann boundary conditions, as explained in Sec.~\ref{sec2}.

While we discussed above rudiments of (classical) field theory on network histories in a covariant framework, 
entanglement measures require a canonical treatment. Locality is made manifest by Hamiltonian 
densities $\mathcal{H}=\tfrac{1}{2}\pi_e^2 + V_\mathrm{eff}(\phi_e)$ on momentum phase-spaces
$T^*\mathcal{C}_e$ consisting of points $(\pi_e,\phi_e)$, with $\pi_e$ denoting 
the momentum conjugated to the configuration field $\phi_e$  
in each fiber over the edge $e$ under consideration. 
The finite resolution structure can be introduced by mimicking smearing prescriptions as follows.
Let us assume there is a short-distance regulator $a$ given by the experiment in question.
Set $\mathrm{grad} \, \phi_e=a^{-1}(T_a-\mathrm{id}_{\mathcal{C}_e})\phi_e$, where 
$T_a$ denotes a spatial translation by $a$. The above equality 
is exact in the sense of Newtonian calculus for infinitesimal quantities 
and its deviation from the limit are well below the experimental resolution limit. 
Furthermore, let $\mathcal{R}$ be a locally finite collection 
of countably many points in the interior of the edge
and denote by $\mathcal{X}_\mathcal{R}(\cdot)$ its indicator function. 
The Hamiltonian adapted to this specific finite resolution structure is simply given by
\begin{eqnarray}
    H_\mathcal{R} := \int_e \mathcal{X}_\mathcal{R} \cdot \mathcal{H}
    \; ,
\end{eqnarray}
which needs to be supplemented by appropriate boundary conditions. 

In the remainder of this article 
we demonstrate the usefulness of quantum networks
for entanglement diagnostics. For simplicity we 
consider adapted networks in Minkowski spacetime 
and use them to evaluate entanglement properties 
of free quantum fluctuations 
in the Poincar\'{e} invariant vacuum. 

In this chapter, we consider the general network depicted in Fig.~\ref{meshGraph}, but focusing at first on its three basic building blocks: A single edge, a loop and the special case of a star graph 
consisting of three edges joined together at a single vertex. 

\subsection{Entanglement diagnostics in $\mathbb{R}_{1,1}$}
\label{subsecEE1p1}
We choose a finite resolution structure characterized by the set
$\{a,\cdots,Na\}$, with $a=const. \in \mathbb{R}$, $N \in \mathbb{N}$, on each edge of the network. $L=Na$ is the long-distance scale and length of the edge.
Introducing the inner product 
$\langle F_1,F_2\rangle_e$ of 
phase space functions $F_1$ and $F_2$ 
by integrating $\mathcal{X}_\mathcal{R} \cdot F_1\cdot F_2$ over the edge, 
the Hamiltonian for a free quantum field $\phi_e$ of mass $\mu$
confined to a single edge 
is given by 
(after appropriate rescaling)
\begin{eqnarray}
\label{hamsinged}
    H_\mathcal{R}
    =
    \frac{1}{2a}
    \big(
        \left\langle \pi, \pi\right\rangle_e
        +
        \left\langle \phi, \mathcal{K}(\phi)\right\rangle_e
    \big) \; ,
\end{eqnarray}
where $\mathcal{K}$ is a bilocal functional 
which is represented by the $(N\times N)$-matrix $K$
relative to the chosen finite resolution structure,
\begin{eqnarray}
    K
    =
    M^2 E
    + \Delta_\mathrm{l}
    \left(
        K
    \right)
    + \Delta_\mathrm{u}(K).
    \label{formK}
\end{eqnarray}
Here, $E$ denotes the unit matrix, $M^2 := \tfrac{1}{N}\mathrm{tr}(K)=2+\mu^2 a^2$, $\mathrm{tr}(K)$ is 
the trace of $K$, 
$\Delta_\mathrm{l}(K)$
and $\Delta_\mathrm{u}(K)$ are the lower and upper triangular
submatrices of the matrix $K-M^2E$, respectively, 
given by  
$
(\Delta_\mathrm{l}(K))_{qs}
=
-\theta(N-s-1/2) \delta_{q,s+1}
$
and 
$
(\Delta_\mathrm{u}(K))_{qs}
=
-\theta(N-q-1/2) \delta_{q+1,s}
$.

At first, we want to study entanglement properties along one single edge of the network only. To this aim, we consider a single edge like the one shown in red in Fig.~\ref{meshGraph} and impose Dirichlet boundary conditions at its endpoints. In this way, we decouple it from the network and we can consider it independently.

In the case at hand, the lower triangular submatrix can be transformed 
into the upper one and vice versa by exchanging rows and columns.
The indices give the multiples of $a$ along the edge $e$, so $\phi_{(e, q)} := \phi_e(q a)$. It is convenient to introduce 
$\bm{\phi}_e:= \phi_{(e, s)} \mathbf{e}_s$, where $(\mathbf{e}_s)^q=\delta^{\;q}_s$, 
and $\mathbf{e}_s \cdot \mathbf{e}_s=\delta_{qs}$.
In greater detail, omitting in this subsection the label $e$ for the specific edge, relative to the finite resolution structure,
\begin{eqnarray}
    H_\mathcal{R}
    =
   \frac{1}{2a}
   \left(
        \bm{\pi}\cdot \bm{\pi} + \bm{\phi}\cdot \left(K \bm{\phi} \right)
   \right)
   \; .
   \label{HReg}
\end{eqnarray}
We can find a unitary transformation $U$ from $\bm{\phi}$ to $\tilde{\bm{\phi}}$
which induces a similarity transformation on $\Omega:=\sqrt{K}$ that diagonalizes it. 
The ground state $\Psi = \otimes_{q\in\lceil N\rfloor}\Psi_q$
of the system relative to the finite resolution structure
is given by the wave function,
\begin{eqnarray}
    \Psi(\bm{\phi})
    =
    \left(
        \mathrm{det}\left(\tfrac{\Omega}{\pi a^2}\right)
    \right)^{1/4}
    \exp{\left(-\tfrac{1}{2} \bm{\phi} \cdot \left(\Omega \bm{\phi}\right)\right)}
    \; .
\end{eqnarray}

In order to probe entanglement properties along the edge $e$, we split 
it into two parts $\mathrm{int}(e)$ and $\mathrm{ext}(e)$,
which is concomitant with dividing the original system 
(mapped onto the finite resolution structure) into a subsystem referred to as interior (I) and 
a subsystem referred to as exterior (E). We decompose $\Omega$ accordingly \cite{srednicki1993entropy, PhysRevD.34.373}, 
\begin{eqnarray}
\label{decomOm}
    \Omega 
    = 
    \begin{pmatrix}
        \Omega_\mathrm{II} & \Omega_\mathrm{IE}\\ 
        \Omega_\mathrm{EI} & \Omega_\mathrm{EE}
    \end{pmatrix}
    ,
\end{eqnarray} %
and similarly for $\Omega^{-1}$.
We choose to compute the reduced density matrix $\rho_\mathrm{E}$ corresponding to the exterior subsystem, that is we integrate out the degrees of freedom 
localized in the interior subsystem giving, 
\begin{equation}
   \rho_{{\mathrm{E}}} \left(\bm{\phi},\bm{\phi^\prime}\right) \sim 
    \exp{\left(-\tfrac{1}{2}\left(
        \bm{\phi}\cdot \left(\gamma \bm{\phi}\right)
        + \bm{\phi^\prime} \cdot \left(\gamma \bm{\phi^\prime}\right)\right)
        + \bm{\phi^\prime} \cdot \left(\beta \bm{\phi}\right)
    \right)}
    ,
\end{equation}
where $\bm{\phi}$, $\bm{\phi'}$ now refer to the exterior collection
of configuration variables relative to the finite resolution structure. Furthermore, 
$\beta := \tfrac{1}{2}\Omega_\mathrm{IE}\, 
{(\Omega_{\mathrm{II}})}^{-1} \, (\Omega_\mathrm{IE})^{\mathrm{T}}$, 
and $\gamma := \Omega_{\mathrm{EE}}-\beta$.

In order to compute the eigenvalues of the reduced density matrix 
$\rho_{\mathrm{E}}$, we need to diagonalize it, which requires 
two more transformations. First, an orthogonal transformation {$V$} 
of the configuration variables, 
$\bm{\phi} \mapsto \gamma^{1/2}_\mathrm{diag} (V \bm{\phi})$ so that
$\gamma =: V^\mathrm{T} \gamma_\mathrm{diag} V$, where $\gamma_\mathrm{diag}$
is diagonal, and subsequently another orthogonal transformation $S$
of the new configuration variables that diagonalizes 
$\Lambda := 
\gamma_\mathrm{diag}^{-1/2} V \beta V^\mathrm{T} \gamma_\mathrm{diag}^{-1/2}$, 
so that $\Lambda =: S^\mathrm{T} \Lambda_\mathrm{diag} S$.
For ease of notation we rename the transformed configuration variables 
by their old names. Then 
\begin{eqnarray}
\rho_{\mathrm{E}}\left(\bm{\phi},\bm{\phi^\prime}\right)
    &\sim&
    \prod_{\phi_q,\phi'_q \in \mathcal{C}_{\mathrm{ext}(e)}}
    \rho_{\mathrm{E}}\left(\phi_q,\phi_q \hspace*{-0.1cm}^{\prime}\right)
    \; ,
    \nonumber \\
    \rho_{\mathrm{E}}\left(\phi_q,\phi_q \hspace*{-0.1cm}^{\prime}\right)
    &=& 
    \exp{\left(-\tfrac{1}{2}
    \left(\phi_q^2 + \phi_q \hspace*{-0.1cm}^{\prime 2}\right)
    +\lambda_q \phi_q \phi_q \hspace*{-0.1cm}^{\prime}
    \right)} \; ,
\end{eqnarray}

where $\lambda_q$ is an eigenvalue of $\Lambda$. Given the eigenvalues of each $\rho_{\mathrm{E}}\left(\phi_q,\phi_q \hspace*{-0.1cm}^{\prime} \right)$ as $p_{n_q}=(1-\tilde{\xi}_q)\tilde{\xi}_q^{n_q}$, with $n_q \in \mathbb Z$ and 
\begin{eqnarray}
    \tilde{\xi}_q
    =
    \frac{\lambda_q}{1+\sqrt{1-\lambda_q^2}}
    \; ,
\end{eqnarray} we can write the spectrum of $\rho_{\mathrm{E}}\left(\bm{\phi},\bm{\phi^\prime}\right)$ as
\begin{equation}
p_{n_{n+1}, \dots, n_{N}} =\prod_{s \in D} \left(1-\tilde{\xi}_s\right)\tilde{\xi}_s^{n_s}  \end{equation}
where we defined $D:=\{q\in \lceil N \rfloor: qa\in\mathrm{ext}(e):=\left[(n+1)a, \dots, Na\right], n \in \mathbb N \}$.
In this language, the total entropy $S$ of entanglement for 
massive quantum fluctuations in the ground state 
confined to a single edge is given by
\begin{eqnarray}
\label{entrN}
    S\left(\bm{\tilde{\xi}}\right)
    &=&
    \sum\limits_{s\in D} S_s\left(\tilde{\xi}_s\right)
    \; ,
   \nonumber \\
    S_s\left(\tilde{\xi}_s\right)
    &=& 
    -\mathrm{ln}\left(1-\tilde{\xi}_s\right)
    -\frac{\tilde{\xi}_s}{1-\tilde{\xi}_s} \; \mathrm{ln}\left(\tilde{\xi}_s\right)
    \; ,
\end{eqnarray}
where again $D:=\{q\in \lceil N \rfloor: qa\in\mathrm{ext}(e)\}$.

For now we are only interested in the dependence of the entanglement entropy \eqref{entrN}
on the set $D$, or referring to Fig.~\ref{meshGraph}, we are interested in its dependence 
on the radius $R$ of the entangling sphere. In fact, as shown in Fig.~\ref{meshGraph}, the entangling sphere radius defines the splitting of the red edge in two intervals; int(\textit{e}) residing inside the entangling sphere, and ext(\textit{e}) residing in its exterior. Hence, by tracing out a sphere of radius $R$ we in fact define the traced out interval int(\textit{e}) on the edge with respect to which \eqref{entrN} is computed. 

It is convenient to normalize the entanglement 
entropy relative to its value for a radius $R$ for which int(\textit{e}) equals half the long-distance scale $L$ and for
quantum fluctuations 
with masses $\mu$ so that $\mu a=10^{-1}$ for a short-distance scale $a$ provided by 
the experimental setup;
$\bar{S}(R) = {S(R)}/{S(L/2)}$, where $R$ is measured in multiples of $L$.

\begin{figure}[t]
	\centering
	\includegraphics{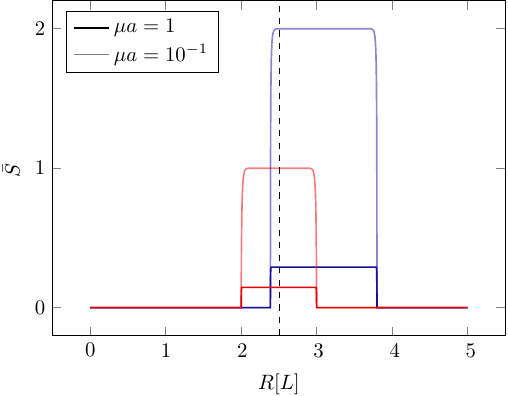}
	\caption{\label{loopgraph} 
        Entanglement entropy of quantum fluctuations confined to the basic 
        building blocks of the mesh graph shown 
        in Fig.~\ref{meshGraph} in terms of the radius of an entangling sphere. 
        In particular, the entanglement entropy is given for the red edge and the blue loop 
        equipped with a finite resolution structure consisting of
        $N=300$ locations for
        two different choices of $\mu a$. 
        The radius of the entangling sphere shown in Fig.~\ref{meshGraph}
        is indicated with a vertical dashed line.
        }
\end{figure}

We developed a code \cite{ourcode} for computing the entanglement entropy 
of quantum fluctuations
on networks that are intersected by entangling surfaces. 
For the single red edge shown in Fig.~\ref{meshGraph},
the results of a numerical computation of $\bar{S}$ are presented in 
Fig.~\ref{loopgraph}.
Referring to Fig.~\ref{meshGraph}, the single edge intersects the 
entangling sphere once. For radii $R<2L$, 
the edge would not intersect the entangling sphere and the entanglement entropy 
for quantum fluctuations localized on the edge would be zero. Similarly the 
entanglement entropy vanishes for radii $R>3L$ since the edge now resides
completely within the entangling sphere.
Furthermore, for radii $R\in(2L,3L)$, the entanglement entropy is independent 
of the finite resolution structure specified by the network. In other words, 
the entanglement entropy is independent of the number of $\phi_q$
residing inside the entangling sphere, and hence of the size of int(\textit{e}) and consequently of $R$.
Intuitively this result can be explained 
as follows. The cross section of a single edge with the surface of the entangling 
sphere is a single point. Any communication channel between vacuum fluctuations
residing inside and outside the entangling sphere has to pass through this cross section.
Equivalently, any entanglement can only be established through this point and this is the case independently of the area of the entangling sphere.
In addition, the entanglement of fluctuations confined to a single edge 
is independent of the angle by which the edge pierces the entangling surface, 
because the fluctuations only perceive the (intrinsic) geometry induced on the edge by 
the embedding Minkowski spacetime. 
Comparing the two parameter choices $\mu a = 10^{-1}$ and $\mu a = 1$ we find that
the entanglement entropy for the former choice is increased relative to the latter.
Intuitively, the two-point correlations between quantum fluctuations confined to the edge
decay exponentially like $K_0(\mu\|x\|)$, where $K_0$
is a modified Bessel function of the second kind, and $\|x\|$ denotes 
the Minkowski distance between the two locations.  Since both locations 
are on the same hypersurface, only their spatial distance $r$ along the edge enters, 
so correlations decay as $(\mu r)^{-1/2} \exp{(-\mu r)}$ for spatial locations 
separated by distances sufficiently larger than the characteristic correlation 
length $\xi=1/\mu$. Hence, for $\mu a =10^{-1}$ the two-point correlations across the entangling sphere decay more slowly than in the case $\mu a =1$, resulting in more vacuum fluctuations 
that are correlated across the surface which leads to more entanglement and consequently in an increase of the entanglement entropy. 
Note that $\mu a = 1$ is an extreme choice bordering at the 
domain of validity of the effective theory describing the quantum fluctuations (see the Appendix).

The finite resolution structure introduces short- and long-distance 
scales, $a$ and $L$, respectively, that give rise to a wavelength window 
characterizing fluctuations whose contributions to the 
entanglement properties of the ground state can be taken into 
account at the operational level. In fact, this window translates 
to a simple hierarchy of length scales $a\ll 1/\mu\ll \mathrm{min}(l,L-l)\ll L$, 
where $l$ denotes the distance from the piercing point to the endpoint of the edge 
residing inside the entangling sphere. 
For characteristic correlation lengths in this window, finite-edge size effects are supported only close to the endpoints of the edge.
For $L\mu\gg 1$ an analytical expression for the total entanglement entropy 
has been found in \cite{Calabrese_2004};  
$S^\mathrm{plat}=-\tfrac{1}{6} \mathrm{ln}(\mu a)$ at $l=L/2$, 
which agrees with our result for the choice of parameters given by
$\mu a = 10^{-1}$.

We now turn to the loop depicted in blue in Fig.~\ref{meshGraph}
which is another basic building element of a mesh graph. As for the case of the red edge, in this section we want to consider entanglement properties along the blue loop as a graph configuration independent of the network. Hence, the four corners of the loop in Fig.~\ref{meshGraph} are to be thought as coupled with Kirchhoff-Neumann conditions for $n=2$ and not for $n=4$ as will be the case when we will study entanglement properties on the whole network.
Relative to a finite resolution structure characterized by the 
set $\{a,\cdots,N a\}$ for each edge, 
such a loop, embedded in Minkowski spacetime, can be replaced 
by a single edge with its endpoints topologically 
identified and equipped with a finite resolution structure 
$\{a,\cdots,4Na\}$. This topological identification can be 
made manifest in the bilocal functional $\mathcal{K}$
represented by the matrix $K$ simply by adding 
terms identifying the endpoints of the extended edge
as nearest neighbors. 
Comparing with the single-edge element discussed above, 
we find that the entanglement entropy for fluctuations confined to 
a loop which pierces the entangling sphere at two different locations 
is twice the value computed for the single edge.
This result is corroborated by an analytical computation.
Let $l_1$ and $l_2$ be the distances of the two edges from 
the locations where they pierce the entangling sphere
to their respective endpoints residing inside the sphere. 
Thus, the intersection of loop and entangling sphere 
is a simple polygonal chain $P$ of length $l_1 + l_2$.
Let us first consider the following hierarchy of 
length scales; 
$a\ll 1/\mu \ll \min(l_1,L-l_1,l_2,L_2-l_2)$.
Under the spell of this hierarchy, 
fluctuations confined to the line segments 
of $P$ are only entangled with fluctuations 
on those edges containing said line segments.
In other words, the above hierarchy effectively reduces 
the loop to two decoupled edges each piercing the entangling 
sphere at different locations. 
We obtain $S^\mathrm{plat}=-\tfrac{1}{6}\mathrm{ln}(\mu a)$
for each of these decoupled edges containing a line segment 
of $P$. The above hierarchy requires  
$a\ll 1/\mu \ll \min(l_s,L-l_s)$ separately for $s\in\{1,2\}$. 
Hence the single-edge case is fully recovered, just doubled. 
The total entanglement entropy for fluctuations confined to the loop 
is $S^\mathrm{loop} = 2\times S^\mathrm{plat}$ which was the assertion. An analogous investigation with an equivalent outcome can be performed with a single edge which is crossed two times by the entangling sphere surface.

This result can be generalized.
Consider fields confined to an arbitrary network having a nonempty intersection
with an entangling shape. 
Focus on the subset of this intersection consisting of all simple polygonal chains $P_n$
with $\mathcal{A}_n$ line segments belonging to edges 
which each pierce the surface of the entangling shape at a different location. 
The total length of $P_n$ is ${}_1l_n+\cdots + {}_{\mathcal{A}_n}l_n$.
Impose the following hierarchy of distance scales associated with each chain $P_n$; 
$a\ll 1/\mu \ll \min({}_1l_n, {}_1L_n-{}_1l_n, 
\cdots , {}_{\mathcal{A}_n}l_n, {}_{\mathcal{A}_n}L_n-{}_{\mathcal{A}_n}l_n)$.
The total entanglement entropy $S^\mathrm{tot}$ of this configuration 
is the sum of each entanglement entropy $\mathcal{A}_n\times S^\mathrm{plat}$  related to each of $N$ polygonal chains assumed to
satisfy the above requirements, 
by a straight forward generalization 
of the above argument in the case of two decoupled edges piercing the surface. 
Let $\mathcal{A}$ denote 
the total number of piercings of the entangling shape, then 
$S^\mathrm{tot}=\mathcal{A}\times S^\mathrm{plat}$.

We close this subsection by remarking that our numerical investigations 
do not require the above decoupling hierarchy of distance scales. 
It is just useful to highlight the universal scaling of the entanglement entropy 
with the area consisting of piercing points of the entangling shape,
as well as to consider a system configuration (including the hardware)
that allows for an extensive entanglement entropy.

\subsection{Entanglement on minimal networks}
An elementary nontrivial network is a so-called star graph 
$\mathcal{N}_\star=(N,E,\iota_*)$
consisting of $|E|$ edges joining a single vertex. 
In other words, $\iota_*(e_\alpha) = (n_\alpha, v)$ for any 
edge $e_\alpha\in E$, where $v$ denotes the vertex 
common to all pairs $\iota_*(e_\alpha)$ and 
$n_\alpha\, (\alpha\in \lceil |E| \rfloor)$ denotes the free endpoint node.   
It is 
another basic building block of a meshlike graph, shown in 
Fig.~\ref{meshGraph} 
as the subgraph depicted in solid 
green lines for the minimal configuration with $|E|=3$.
There are two type of boundary conditions 
relevant to the analysis of physical processes 
confined to such a minimal network; 
the boundary conditions at the free endpoints of each edge 
and the boundary conditions at the single vertex connecting
all edges. A natural choice for the former are Dirichlet 
boundary conditions in accordance with the requirement 
that all fields are confined to the network \cite{berkolaiko2012introduction}. At the vertex
we impose Kirchhoff-Neumann coupling conditions which 
generalize Kirchhoff's circuit laws.
As discussed in Sec.~\ref{secII}, these conditions are essential for guaranteeing the conservation of probability current and ensuring the smoothness of the quantum field across vertices.

For simplicity, we equip every edge with the same 
finite resolution structure $\mathcal{R}$ characterized by the set 
$\lceil N\rfloor a$ and denote the representation 
of the Laplace operator relative to this structure 
by $\Delta_\mathcal{R}$. Relative to the chosen finite resolution structure, solving  Eq.~\eqref{bdyCondGen} with $A$ and $B$ given in \eqref{ABkn} with $n=3$ leads to \cite{besse2021numerical};
$
    \phi_v=\tfrac{1}{N}
    \sum_{\alpha\in\lceil |E|\rfloor} \phi_{(\alpha,1)}.
$ The Hamiltonian of 
ground state fluctuations confined to a general 
star graph in Minkowski spacetime is then given by 
$H_*=|E| H_\mathcal{R} + C_\mathcal{R}$, where $H_\mathcal{R}$ is 
given in (\ref{hamsinged}), and $C_\mathcal{R}$ denotes 
the energy shift relative to the decoupled network configuration
considered above, 
$C_\mathcal{R}=\tfrac{1}{2a} \tfrac{1}{|E|}
(\sum_{\alpha\in\lceil |E|\rfloor} \phi_{(\alpha,1)})^2$, 
where the configuration space is now parameterized by the 
set $E$ and the finite resolution structure $\mathcal{R}$. 
Explicitly, the index pair $(\alpha,1)$ refers to 
an event located $1 a$ away from the vertex location on 
$e_\alpha$. 

\begin{figure}[t]
	\centering
	\includegraphics[width=0.48\textwidth]{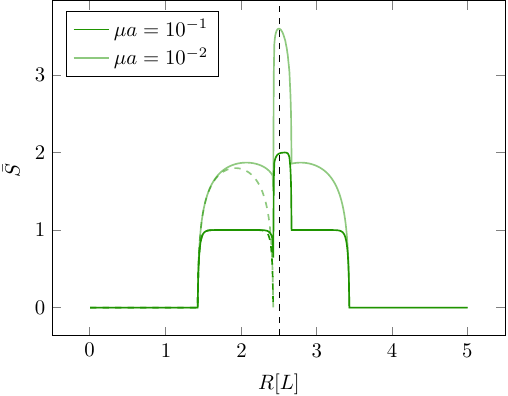}
	\caption{\label{3edgesvertex2} 
    Entanglement entropy of vacuum fluctuations confined
    on a network idealized as a minimal star graph with three 
    edges joined at a single vertex (referring to Fig.~\ref{meshGraph})
    in terms of the entangling sphere radius
    for two different values of $\mu a$.  
    For comparison the entanglement entropy of vacuum fluctuations 
    confined to a single edge with Dirichlet boundary conditions 
    at both its endpoints is shown in dashed lines
    for both cases.} 
\end{figure}

As a first example we consider a network idealized as a minimal star graph with three edges joined at a single vertex in Minkowski spacetime such as the one displayed in Fig.~\ref{meshGraph} in green with solid lines.
Notice that the coupling term $C_\mathcal{R}$ in the Hamiltonian introduces additional terms in $K$ between $\phi_{(1,1)}$, $\phi_{(2,1)}$ and $\phi_{(3,1)}$ with respect to the uncoupled case.
Implementing these conditions in \eqref{formK} and into our code \cite{ourcode} we can numerically investigate the entanglement entropy of the quantum field on the network. 
In Fig.~\ref{3edgesvertex2}, 
the entanglement entropy of vacuum fluctuations is shown 
as a function of the entangling sphere radius $R$ and for different 
values of $\mu a$.
With increasing $R\in (1.4L,2.4L)$ an increasing number of fluctuations 
localized (in accordance with the finite resolution structure)
on the edge partially residing inside the entangling sphere 
are traced out. At $R=2.4 L$ the only vertex of this minimal
network idealization crosses the entangling sphere.
Increasing the radius further, $R\in (2.4L,2.7L)$, 
parts of the other two edges reside inside the entangling sphere.
As a result we have now two channels that communicate the 
vacuum entanglement between the interior and the exterior 
of the entangling sphere. 
For $R\in (2.7L,3.4L)$ a second edge in Fig.~\ref{meshGraph}
resides completely inside the entangling sphere, leaving 
only parts of the third edge in the exterior. 
This configuration is similar to the one considered before 
for $R\in(1.4L,2.4L)$ where only parts of a single edge resided 
in the interior of the entangling sphere, while the other two 
edges resided completely in the exterior. 

Decreasing the value of $\mu a$ increases the entanglement 
of vacuum fluctuations for reasons we already explained in 
Sec.~\ref{subsecEE1p1}.
Choosing $\mu a = 10^{-1}$ and 
either $R\in (1.4L,2.4L)$ or $R\in (2.7L,3.4L)$
we obtain the same value for the entanglement entropy 
as in the case of vacuum fluctuations confined to 
a single edge $e$: $\bar S= \bar S_e=1$. 
On the other hand, for $R\in(2.4L,2.7L)$, when 
the network intersects the surface of the entangling sphere 
twice, we find indeed $\bar S=2 \times \bar{S}_e$.

The case $R=2.4L$ deserves further discussion. For this radius
the surface of the entangling sphere intersects the vertex
and, as a consequence, 
the entanglement entropy shows an enhanced sensitivity 
on the Kirchhoff-Neumann boundary conditions 
leading to a decrease in the entanglement entropy. 
The reason for this behavior is a decrease in the strength 
of correlations across the vertex since the coupling 
$C_\mathcal{R}$ term is suppressed by a factor $1/|E|$ 
relative to the usual nearest neighbor coupling. 
Furthermore, note that $L$ is chosen so that $L/2 \gg 1/\mu$ and 
we are still analyzing the case $\mu a=10^{-1}$.
This guarantees a plateau of the entanglement entropy
for values of $R$ close to $R=2L$ and $R=3L$.

These numerical experiments can be used to answer 
an obvious question: Can the entanglement diagnostics 
employed in this work be used to characterize the 
underlying network infrastructure? The answer is in 
general negative. For instance, consider the case
$R\in(1.4L,2.4L)$ in the above experiment, that is, 
a minimal network where the interior of only one edge is intersected 
once by an entangling surface. In this case the entanglement 
entropy of vacuum fluctuations confined to this network 
equals the entanglement entropy measured on a single edge 
(provided, of course, its interior is intersected once 
by the entangling surface, as well). As a consequence, 
both network structures cannot be distinguished based 
on the performed entanglement diagnostics.

\subsection{Infrastructure and entanglement}
\label{sec3B}
Comparing the entanglement entropy of vacuum fluctuations 
confined to a minimal network with the corresponding 
quantum information measure on a single edge, both reported in
Fig.~\ref{3edgesvertex2}, we observe deviations from 
the single-edge configuration as the entangling sphere 
approaches the vertex of the minimal star graph. 
These deviations are triggered by correlations 
reaching from the interior edge
across the vertex to the two edges residing outside the 
entangling sphere for both values of $\mu a$.
As the entangling sphere is extended towards the vertex, 
the entanglement entropy of the minimal network increases 
relative to the single-edge case, simply because more 
localized fluctuations per edge contribute to the 
entanglement. For lower values of $\mu a$ the effect 
is more pronounced due to the exponential decay of spatial correlations 
with $\mu a$ discussed above. In other words, 
a lower value of $\mu a$ implies a longer correlation length 
and, thus, an extended entanglement on larger network scales 
involving an increasing number of spatially separated fluctuations. 

For $\mu a=10^{-1}$ the entanglement entropy for the 
single-edge configuration develop a plateau which represents 
an upper bound on the entanglement entropy for the minimal 
network even as the entangling sphere approaches the vertex
which opens up more communication channels. So despite the 
increase of correlations between the interior and the exterior of the 
entangling sphere, 
the entanglement entropy never exceeds the plateau value 
of the single-edge configuration. 
This shows how the Kirchhoff-Neumann junction conditions 
control the impact of the vertex on the entanglement;
the vertex decreases the strength of correlations 
across the junction by a factor $|E|^{-1}$ relative 
to the correlation strength between fluctuations 
localized sufficiently far away from it. 
Note that the entanglement entropy does, however, not 
vanish when the vertex intersects the surface of the entangling sphere.

A third effect of the network infrastructure on the entanglement 
of vacuum fluctuations confined to it concerns the presence 
of loops and is analyzed in our code \cite{ourcode}, as well. 
Consider again the minimal network idealized as a star graph 
with three edges joining a single vertex, but now we connect 
two free endpoints and form a loop that is joined at the vertex 
by the single remaining edge, as indicated in Fig.~\ref{meshGraph}
with the extension represented by the open polygon shown in green dashed lines. 
Loops in the infrastructure can counteract the impact of the vertex 
on the entanglement of vacuum fluctuations confined to the network.
More precisely, if the entangling surface is close to the vertex, 
the presence of a loop can increase the entanglement 
entropy relative to the corresponding minimal graph configuration, 
provided the size of the loop is smaller than the 
typical correlation length. Then, loosely writing, the loop admits 
additional correlations between localized fluctuations inside and 
outside the entangling surface, respectively.
If the loop size is larger than the typical correlation length, 
then, relative to the quantum information measure we use,  
the network configuration is effectively equivalent to the 
simple minimal network. 

In conclusion, the entanglement entropy of vacuum fluctuations localized on
more complicated networks constructed by joining several minimal networks 
is determined by the interplay of all the aforementioned effects.
The following rule of thumb reflects the intuition supported by our numerical experiments. 
Each new edge intersecting the entangling surface adds more vacuum fluctuations 
that are correlated across the surface which leads to more entanglement, 
each new vertex sufficiently close to the entangling surface decreases the 
strength of correlations between vacuum fluctuations localized on 
different edges joined by this vertex, and each new loop 
of sufficiently small size
(relative to the typical correlation length)
further entangles the interior and exterior of the surface. 

As a result, if the typical correlation length is large compared to the maximal 
size of any edge in a subgraph nested inside an idealized network, then 
the entanglement entropy of vacuum fluctuations measured by this
part of the infrastructure will depend on its topology in addition 
to the number of its intersection with the entangling surface.

\label{sec3}
\section{Area Scaling on Networks} 
\label{eeGraph}
The results of the previous section on the entanglement 
entropy of vacuum fluctuations confined to a minimal network -- 
idealized as a star graph -- which intersects a given entangling surface, 
can be used to deduce some entanglement properties
of vacuum fluctuations
on more sophisticated networks.
Consider an arbitrary network, for instance the 
one depicted in gray in Fig.~\ref{meshGraph}, 
and its intersection with the entangling surface (in this 
case a sphere).
At each intersection point, the typical correlation length scale determines a subpart of the graph, containing the intersection point, which we refer to as a {\it local subgraph}. A local subgraph either 
contains at least one vertex (connected to the intersection point), 
or it coincides with an entire edge or parts of it. 
If the correlation length is such that each local subgraph 
is disconnected from all the others (located at different 
intersection points), then, by our results from the previous section, 
we conclude right away that the entanglement entropy 
becomes, effectively, an extensive quantity relative to 
the disconnected local subgraphs, 
\begin{eqnarray}
\label{sumEE}
    S
    =
    \sum_{a\in \lceil n_\mathrm{s}\rfloor}
    S_a
    \; ,
\end{eqnarray}
where $n_\mathrm{s}$ denotes the number of local subgraphs. 
The formula (\ref{sumEE}) for the total entanglement entropy
holds, in particular, in the following situation.
Suppose the correlation length satisfies the hierarchy
$1/\mu \ll \mathrm{min}(l_a,L_a-l_a)$ for the edges 
in each subgraph on which the intersection points are located. 
Under the spell of this hierarchy the subgraphs effectively reduce to these edges when considering the entanglement entropy of a field
with mass $\mu$, in accordance with the results 
of Sec.~\ref{subsecEE1p1}. 
In this case all subgraphs become equivalent and the 
entanglement entropy is given by $S=n_\mathrm{s}\times S_1$. 
Again, this formula generalizes the idea of 
Sec.~\ref{subsecEE1p1} to arbitrary networks upon identifying 
a single edge (on which the intersection point is located) with a
disconnected subgraph.
If the above hierarchy of length scale is inverted, then it is possible 
that vacuum fluctuations located in the neighborhoods of different 
intersection points become correlated. In this case the entanglement 
entropy ceases to be an effectively extensive quantity and increases
relative to the previous case.

We can elaborate the above findings further:
If the typical correlation length of vacuum fluctuations confined 
to a network is smaller than the length scales characterizing 
a given set of effectively disconnected (relative to the correlation length)
subgraphs, each of which consists of connected components 
containing a single intersection point, then 
the entanglement entropy is an extensive quantity with 
respect to the disconnected subgraphs. 
Intuitively, the fluctuations cannot resolve the infrastructure 
beyond the individual subgraphs. In the extreme case that only 
the sizes of the subsystems on the intersected edge can be resolved, 
the entanglement entropy becomes maximally extensive in the sense 
that there is no smaller subgraph relative to which this property 
can hold. 
Relative to the same set of subgraphs, if the correlation length 
is increased, the vacuum fluctuation start to probe an increasing 
amount of facets of the network. At an intermediary step it might be 
possible to identify a new set of effectively disconnected subgraphs 
and establish an extensive entanglement entropy on larger scales. 
Relative to the extreme case discussed above, however, 
the entanglement entropy on each subgraph will depend on 
a complicated interplay of the effects discussed in Sec.~\ref{sec3B}, 
albeit the characterization as an extensive quantity 
relative to the subgraph level is valid. 
It is exactly in this regime that we leave the usual $(1+1)$-dimensional QFT treatment in favor of a full quantum graph description.

\subsection{Networks and spacetime entanglement}
An exciting application of networks is as diagnostic devices 
to probe physical phenomena (compactly) supported in the 
embedding spacetime by solely employing fields 
confined to the networks and their {\it two-dimensional} 
piecewise smooth Lorentzian histories. 
This requires networks idealized by meshlike graphs with compact 
extensions in all directions tangent to the instantaneous hypersurfaces 
in which they are at rest. Such an infrastructure naturally offers two 
types of investigations. The first concerns the similarity between
physical systems confined to given networks and their counterparts 
enjoying compact but otherwise unrestricted spacetime support.
Provided the similarity grows beyond experimental uncertainties, 
both systems cannot be discriminated (relative to the experiments), 
and the theory describing the system confined to the network and 
its two-dimensional histories can be considered a sufficient approximation 
of the (possibly ill-defined) continuum theory within the experimental accuracy. 
The second concerns employing networks adapted to probe aspects of the embedding 
spacetime geometry as captured by physical systems confined to the networks.

In field theory entanglement entropy becomes a continuum concept
which is intrinsically dominated by short-distance correlations 
across the entangling surface. If the entangling surface is 
idealized as a border of infinitesimal width, and if it is wrongly
assumed that details concerning a short-distance completion 
are irrelevant, then the entanglement entropy cannot be computed 
in quantum field theory. In other word, entanglement entropy as a 
quantity probing the spacetime continuum requires a short-distance completion. 
This seems to imply that entanglement entropy 
is not a meaningful quantum information measure in field theory 
but this possibility is not enforced since even in the absence 
of a short-distance completion (intrinsic requirement), 
the definition of entanglement entropy can be adapted 
at the level of the infrastructure (extrinsic requirement). 
For instance, it is impossible to construct an infinitesimally 
thin entangling surface. Introducing a physical surface 
amounts necessarily to specifying a minimal distance scale.
This is not done at the intrinsic level, since the surface 
is considered to be part and parcel of the hardware infrastructure 
which remains unresolved in terms of physical degrees of freedom. 
This is an example where any experiment comes automatically 
equipped with a finite resolution structure that we need to take into 
account at the operational level and that poses an
extrinsic but at the same time principal resolution limit. 
So to offer a logical alternative it may be that we have to relax 
the formal definition of observables by taking into account 
finite resolution structure induced by an external object. 
Any measurement of entanglement entropy is performed relative 
to such a structure and this is inevitably unavoidable. 

Our strategy in this section is to pursue the first investigation 
outlined above using entanglement entropy as a quantum information 
measure relative to a finite resolution structure for vacuum fluctuations 
in Minkowski spacetime,  
which simultaneously serves as a reference spacetime 
concerning the second type of investigations, that is, for 
probing generic spacetime geometries in which a network 
is embedded. 

Concretely, we show in this section that a specific 
network class exists which allows to extract 
the entanglement entropy of vacuum fluctuations 
in the embedding Minkowski spacetime, albeit the network 
history is an embedded piecewise smooth 
two-dimensional Lorentzian history. 
This is an important example for emerging 
spacetime properties on lower dimensional structures. 
Let us stress again that the network does not serve
as a discrete structure like a lattice, 
rather it comes equipped with a finite resolution structure. 
The network is a physical manifestation (hardware)
of the support structure on which the physical 
degrees of freedom are bound to exist. 

There are two reasons for a coarse-grained modeling 
of the continuum physics on the network. 
One is invited by the finite resolution structure itself 
and concerns the small-scale description of continuum 
quantities on the network. 
The other is the resolution of the embedding hypersurface 
(or spacetime) in terms of vertices. 
At each point in the interior of the edge there is 
a plane in the hypersurface tangent to the edge at this point.
Only at the vertices can a path along the edge leave 
the local tangent plane in the hypersurface to 
extend in the remaining dimension. 
Thus, the local vertex density can be used as a measure 
for the local filling of the hypersurface by the network.
We choose a unit hypersurface volume to construct the 
local vertex density $d_\mathrm{v}$. 
The typical intrinsic reference scale is given by 
$\xi := 1/\mu$, where $\mu$ denotes the mass parameter
of the vacuum fluctuations under considerations. 
This implies an effective intrinsic density scale for a 
coarse-grained description. Certainly, if the 
vertex density exceeds this scale, that is, if 
$d_\mathrm{v}\gg 1/\xi^3$, then fluctuations
are blessed with ignorance concerning the coarse-grained 
(extrinsic) infrastructure they are confined to and probe 
the full embedding spacetime, subject only to the 
intrinsic resolution scale set by their mass parameter.
Of course, this assumes that the vertex density $d_\mathrm{v}$ 
complies with the isometries of the embedding spacetime; in the
case under consideration, $d_\mathrm{v}$ is assumed to 
be spacetime homogeneous to comply with the Poincaré isometry 
of Minkowski spacetime. 
Provided these two conditions are met, networks approximate 
the embedding spacetime, subject only to the intrinsic resolution limit 
of the fluctuations that are employed as spacetime probes.
We refer to these networks as {\it dense} networks. 

In the following we consider, for simplicity, a three-dimensional 
regular grid of finite size, where, in particular, all edges 
have the same length $L$. This network has one vertex per unit 
volume $L^3$, so $d_\mathrm{v}=1/L^3$. Hence, the above relation 
between the intrinsic and extrinsic resolution scale becomes simply
$L\ll\xi$. 

We stress again that this network configuration must not be 
confused with a lattice, since contrary to the latter 
each of the network components support fluctuations on
a two-dimensional Lorentzian submanifold embedded 
in a four-dimensional Minkowski spacetime. 
In other words, the fluctuations themselves act always as
two-dimensional spacetime probes, contrary to field theory 
on a lattice where each nonempty vertex 
supports fluctuations that are sensitive to all spacetime dimensions. 

For the following numerical experiments we refer the reader to the code 
\cite{ourcode}. Our discussion 
focuses on two regimes: One for which the entanglement entropy
of vacuum fluctuations confined to a grid-like network is an extensive quantity 
relative to isolated subgraphs as discussed above, and the other 
for which the fluctuations probe the network on large scales 
so that the entanglement entropy manifests itself as
an intensive quantity. 

\subsubsection{Effectively extensive entanglement entropy}
The discussion above implies that there is always a mass range of vacuum fluctuations 
for which the entanglement entropy of these fluctuations confined to 
a grid-like network is given by Eq.~\eqref{sumEE} relative 
to a given finite resolution structure. In the extreme case, $\xi \ll \min \{l_e,L-l_e\}$ for each edge $e$ containing a crossing point. This case corresponds effectively to a 
collapse of any choice of subgraphs connected to the entangling surface
to a collection of edges piercing the entangling surface. 
In particular, the vertices cease to be the components controlling 
the spread of entanglement over the network.
In other words, for this hierarchy between the typical 
internal and external distance scales,
the ideal network description is reduced from graphs including vertices 
to just the edges piercing the entangling surface. 
Then, according to Eq.~\eqref{sumEE}, the total 
entanglement entropy becomes effectively an extensive quantity 
given by $S=\mathcal{A}\times S_{1+1}$, where $\mathcal{A}$ 
is again the number of intersection points where the edges 
pierce the entangling surface. 
Due to the embedding of the network into Euclidean hypersurfaces 
geometrical information is available like the angle 
between the local normal of the entangling surface and the piercing edge. 
However, the vacuum fluctuations only probe the intrinsic edge geometry, 
or -- if instead of an instantaneous network configuration a freely falling 
network is considered -- the intrinsic geometry of the piecewise smooth 
Lorentzian network history. \textit{Nota bene}, this is again because the 
fluctuations are not merely restricted but confined to the network 
and therefore cannot experience extrinsic geometry, 
which is why $S_{1+1}$ is independent of this information. 
As a consequence we may assume that all edges intersecting the 
entangling surface are aligned with the local normal, corresponding to a radial configuration in the case 
of an entangling sphere.

In the present subsection the entangling surface is considered to be spherical and thus, for each radius $R$ we obtain
a finite collection of radial edges piercing the surface 
of the entangling sphere at an arbitrary location, 
depending on which network is concretely employed
(or implemented in the numerical experiments). 
This leads, in general, to the number $\mathcal{A}$ 
of piercing points being some
unknown function of the radius $R$.
The analytical approach can substitute numerical experiments 
and is a suitable tool to diagnose, for instance, 
the entanglement properties of fields
confined to freely falling networks in curved spacetimes, 
provided $\mathcal{A}(R)$ is known or can easily be modeled.
Different network configurations 
and/or different shapes of regions containing degrees of freedom 
that we integrate out result in different values for $\mathcal{A}$.

Considering regular three-dimensional grid graphs 
with edges of length $L\ll R$, 
there is approximately one edge piercing the entangling surface
per $L^2 \pi/4$, leading to $\mathcal{A}(R)= 16 (R/L)^2$, for the case of a spherical entangling surface of radius $R$. 
If the entangling surface were cubic instead of spherical, 
with the same area of $4\pi R^2$, the density of piercing edges would 
have been $1/L^2$, yielding a higher number of piercing edges $\mathcal{A}(R)= 4\pi R^2/L^2$.
As we will investigate further below, the entanglement
entropy might exhibit a shape dependence; for example when tracing out cubic and spherical regions with equal surface areas. To be able to give a 
verdict on it, we compare both cases not only with same surface area 
but also with the same density. 
If differences persist, it would indicate a more profound shape dependence of the entanglement entropy.
Therefore, for the sake of the comparison, we already set
the density of the collection of piercing edges to be $1/L^2$, as in the cubic case. We will employ this choice always, regardless of the shape of the entangling 
surface.
The total entanglement entropy for
such a configuration is thus given by $S(R)=\mathcal{A}(R) S_{1+1}(L/2)=4\pi R^2/L^2 S_{1+1}(L/2)$.

\begin{figure}[t]
\centering
	\includegraphics{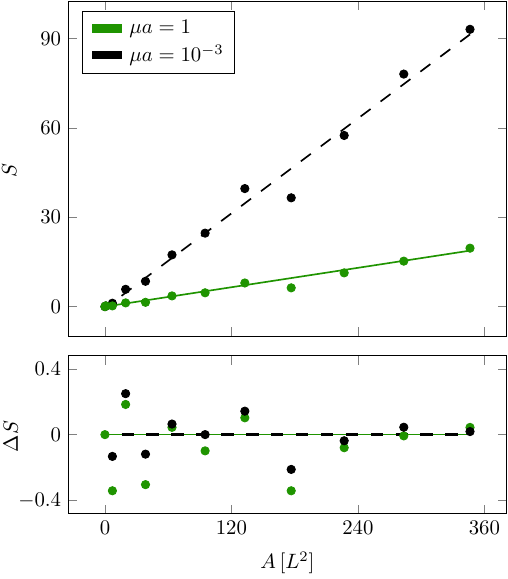}
	\caption{\label{3D} 
    The upper plot shows the entanglement entropy $S_\mathrm{grid}$ of 
    quantum fields in the ground state confined to 
    a three-dimensional 11x11x11 regular grid graph with $11^3$ vertices
    and universal length $L=7a$ of the edges in terms of a 
    minimal distance scale $a$ implied by an assumed 
    finite resolution structure associated with some experiment.  $S_\mathrm{grid}$ is shown in terms of the area $A$ of the entangling sphere.
    Data points correspond to numerical experiments 
    for two different values of $\mu a$. 
    For $\mu a =1$ (displayed in green color), the analytical 
    approximation $S(R)$ (green solid line) described in the main text
    is in good agreement with the data points (from numerical experiments).
    For the case $\mu a =10^{-3}$ (black data points) there is no analytical approximation for the given long-distance scale $L$.
    For convenience, a fit $S_\mathrm{fit}$ (black dashed lines) of the data from 
    the numerical experiments has been added.
    The lower plot shows the quality of the agreement using 
    $\Delta{S} = \delta S - 1$ with 
    $\delta S = S(R)/S_\mathrm{grid}$ for $\mu a =1$
    and $\delta S = S_\mathrm{fit}/S_\mathrm{grid}$ for $\mu a =10^{-3}$.}
\end{figure}

The calculation 
for an individual edge was outlined in Sec.~\ref{subsecEE1p1}. Depending on the mass of the quantum field and on the length of the edges analytic expressions for $S_{1+1}(L/2)$ can be available.
In Fig.~\ref{3D}, in green color, we compare the analytical result $S(R)$, obtained by employing \eqref{finalS1+1} for $S_{1+1}(L/2)$, 
to the numerical entanglement entropy.
Note that numerically the surface area of the entangling sphere is overestimated by a factor of $4/\pi$, which we compensate for by multiplying the numerical
entanglement entropy on these grid graphs with the inverse 
factor $\pi/4$, thus finding agreement to the case of a collection of radial edges with one puncture of the entangling sphere per $L^2$.

The distance hierarchy $\xi=1/\mu \gg L$ lies outside the domain of validity 
of the analytical approach for obvious reasons; 
under the spell of this hierarchy 
entanglement spreads over entire subgraphs of the network.
As a consequence, in order to assess the entanglement 
of such a system, it is insufficient to analyze only 
the immediate neighborhood of an entangling surface.
This hierarchy then demands a numerical investigation 
of entanglement properties, and we employ again 
the code 
\cite{ourcode} for a regular grid graph.
For the results of our numerical experiment 
see Fig.~\ref{3D}, in black color, and the figure caption for further details. 

It is of fundamental importance that area scaling 
emerges in Fig.~\ref{3D} 
for both hierarchies, $\xi=1/\mu \leq L$ and $\xi=1/\mu > L$, 
between the intrinsic and extrinsic distance scale 
characterizing the total system.
This is remarkable, because area scaling is a fingerprint 
for entangled fields extending in full spacetime, while 
we probe entanglement properties of fields confined to 
piecewise-smooth two-dimensional Lorentzian histories
with instantaneous field configurations confined to 
edges of a given network. 

\subsubsection{Emerging spacetime properties of entanglement}
In order to exploit this idea further, we investigate 
whether there is a class of networks serving as support structures 
for fields confined to them 
that allow to probe entanglement properties of the same type of fields 
extending in compact regions of Minkowski spacetime
void of confining networks. 
In other words, we study the potential for spacetime entanglement 
emerging on piecewise smooth $(1+1)$-smooth Lorentzian histories.

As already mentioned, the proportionality factor between the entanglement entropy
and the area of the entangling surface, 
for instance between the entanglement entropy and $4\pi R[L]^2$
shown in Fig.~\ref{3D}, depends on the specific network 
implemented in the (numerical) experiment. 
In fact, in order to obtain the same proportionality as 
for the entanglement entropy in $(1+3)$-dimensional 
Minkowski spacetime \cite{srednicki1993entropy} with a given accuracy, 
the density of the idealized network in Minkowski spacetime 
has to be sufficiently high, entailing eventually 
the limit $L/R\rightarrow 0$ when the support of the fields 
confined to the network approximates a portion 
of the embedding four-dimensional Minkowski spacetime
to arbitrary accuracy. 
This limit, however, cannot be achieved for the following reason:
Before the short-distance cutoff below which 
the effective field theory requires a (partial) completion is reached, 
at the very least we know that our ignorance 
about the small-scale structure of spacetime 
in the semiclassical approximation has to be resolved eventually, 
the minimal distance scale $a$ implied by the finite resolution structure 
of any (numerical) experiment prohibits the required continuum limit. 
This is justified by any experiments, since networks represent 
physical hardware infrastructures that can be manufactured 
only with a finite density. 

Alternatively, higher-dimensional phenomena emerge 
if $L/\xi \ll 1$ and the network density 
exceeds one vertex per correlation length $\xi$ cubed.
In this situation, the correlation length becomes 
too large to resolve the fine-grained network structures, 
that is, the entanglement properties become insensitive 
to the network details, which is why quantum fields
confined to $(1+1)$-dimensional Lorentzian histories 
show an area scaling in their entanglement entropy. 

In general, area scaling of the entanglement entropy 
is inevitable since massive degrees of freedom always allow
to localize correlations in a neighborhood containing the entangling 
surface. If the fields are not confined to a network but extend 
unrestricted to compact regions in spacetime, 
then they carry angular momentum, as well, which combines 
with their intrinsic mass to an effective mass. 
Even if the intrinsic mass vanishes, the angular momentum 
effectively limits entanglement to nearest neighbors 
(relative to some finite resolution structure)
across the entangling surface \cite{srednicki1993entropy}.
Thus, area scaling is inevitable for the entanglement entropy 
of ground state fluctuations in $(1+3)$-dimensional Minkowski spacetime. 

However, fields confined to a network experience the intrinsic geometry of
$(1+1)$-dimensional Lorentzian histories embedded in $(1+3)$-dimensional 
Minkowski spacetime with instantaneous field configurations confined 
to one-dimensional edges. Hence they do not carry angular momentum
and the correlation length is solely determined by their intrinsic mass. 
Therefore, while an area law for $ L \gg \xi$ could be expected because of the extensive property of the entanglement entropy in this regime 
(as seen by the green line in Fig.~\ref{3D}), 
it is very important that area scaling holds for $L \ll \xi$, as well (see the dashed black line in Fig.~\ref{3D}).
This remains true even when
correlations spread deep into the entangling sphere.
In this situation 
correlations between fluctuations localized 
on different edges far apart (relative to the background geometry)
still show area scaling of their entanglement entropy. 
This is only possible if the interplay of effects 
analyzed in Sec.~\ref{sec3B} amounts to simulating 
the presence of angular momentum. 
In other words, we can interpret the results concerning 
the hierarchy $L \ll \xi$ shown in Fig.~\ref{3D}
as an emergence of angular momentum for the fields 
confined to the network which induces area scaling 
for their entanglement entropy. 
Therefore, networks equipped with fields confined to them 
are capable to trace fingerprints of the physics 
of these fields when they extend unrestricted (deconfined)
to higher-dimensional spacetime regions.
Indeed, networks arise as potent arenas where 
phenomena experiencing all spacetime dimensions 
can be investigated using lower-dimensional probes. 

\subsection{Shape dependence}
\begin{figure}[t]
\centering
\includegraphics{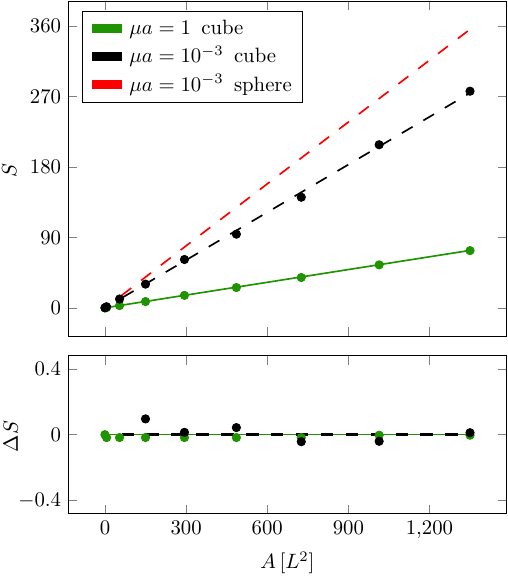}
	\caption{\label{3Dcube} 
    The upper plot shows the entanglement entropy $S_{\rm grid}$
    of quantum fields in the ground state confined to a three-dimensional 17x17x17 regular grid graph with $17^3$ vertices and universal
    length $L=4 a$ of the edges in terms of a minimal distance scale
    $a$ implied by an assumed finite resolution structure associated
    with some experiment. $S_\mathrm{grid}$ is shown in terms of the area $A$ of the entangling cube.
    Data points correspond to numerical
    experiments for two different values of $\mu a$. 
    For $\mu a = 1$ (displayed in green color) 
    the analytical approximation $S(R)$ is shown 
    using a green solid line.  
    A fit $S_\mathrm{fit}$ of the data for $\mu a =10^{-3}$ is depicted
    using a dashed black line.
    For comparison, the fit to the entanglement 
    entropy for an entangling sphere 
    introduced in Fig.~\ref{3D} is shown in red.
    The lower
    plot shows the quality of the agreement using 
    $\Delta {S} = \delta S - 1$
    with $\delta S = S(R)/S_\mathrm{grid}$ for $\mu a = 1$ 
    and $\delta S = S_\mathrm{fit}/S_\mathrm{grid}$ for
    $\mu a = 10^{-3}$.}
\end{figure}

In this subsection we investigate entanglement of quantum graphs 
on the regular grid graph as in the previous subsection, but 
across an entangling surface of a different shape.
Concretely, we trace over degrees of freedom located inside cubic 
regions of different volume,
compute the respective entanglement entropies
using our code \cite{ourcode}
and compare them to the case of entangling spheres. 
The entanglement entropy in terms of the area $A$ of the surface of the cube 
is shown in Fig.~\ref{3Dcube}. 
\label{secgrid}	

This result is a further concrete confirmation 
of the area law.
It is worth noting that implementing an entangling cube 
yields data points scattered closer to a perfect area law, 
as compared to the case of an entangling sphere, 
and the dip observed in Fig.~\ref{3D} for $A = 180 L^2$ 
is no longer present. This is because in the case we trace out 
a spherical region of a regular three-dimensional grid, 
the entangling surface might intersect network vertices.
As explained in Sec.~\ref{sec3B} for the case of a star graph with three edges, 
once the entangling surface resolves network vertices, 
the entanglement entropy is locally decreased 
and in the case of a grid graph even more 
since each vertex joins six edges. 
By tracing out a cubic region of a regular three-dimensional grid network, 
it is possible to choose cube sizes such that vertices of the network 
remain unresolved.

Comparing the entanglement entropy results obtained by tracing out a sphere and a cube in Fig.~\ref{3Dcube} we find the following.
Under the spell of the hierarchy $L \gg \xi$, the numerical experiments 
agree to very good precision on the value of the entanglement entropy $S(R)$
for the same areas of the entangling surfaces (where we note that for the cube the discrepancy to a perfect area law, i.e., $\Delta S$, is smaller compared to the one for the sphere).
In fact, the analytically determined entanglement entropy given as a green line for the cube is identical to the one for the sphere. This is expected since the total entanglement entropy can be described 
by the entanglement entropy of fields confined to a collection of edges 
piercing the entangling shape, with one edge per unit surface area $L^2$. 
Since the entangling sphere and the cube are chosen to have the same area, 
the expectation follows.
For the other hierarchy, $L\ll \xi$, 
the entanglement properties of fields confined to the network
depend on the shape of the entangling surface. 
The reason is that under the spell of this hierarchy
the entanglement is no longer restricted to locations 
within a neighborhood of the entangling surface that
just encompasses single edges across the surface. 
Instead degrees of freedom located deeper in the volume
of the entangling surface get involved in entanglement 
across the surface which leads to the observed 
shape dependence, despite still providing an area scaling. 
In particular, Fig.~\ref{3Dcube} shows that 
the entanglement entropy related to an entangling sphere 
is larger than the one related to an entangling cube 
even if both entangling objects have the same surface area.
It is interesting to speculate whether networks 
can be used to infer the shape of the entangling surface
(not just its area). 
As a result, the network approach shows that, for $L\ll \xi$, the entanglement entropy of vacuum fluctuations confined on the network is not fully determined by the area of the
traced out region, but also by its shape, with information about it encoded in the proportionality factor.

\section{Conclusions and Outlook}

In this article, we presented a novel approach to investigate field theoretical phenomena by employing ideal networks equipped with fields confined to them, as diagnostic tools. As a first application, we have explored entanglement properties 
of quantum fields confined to networks histories embedded in the Minkowski background. 
Our findings show that although the fields are defined on 
$(1+1)$-dimensional Lorentzian histories 
with instantaneous field configurations localized on the 
one-dimensional edges of the network,
the entanglement entropy scales with the \textit{area} of the traced-out region, 
indicating the potential to explore nontrivial properties 
tight to the embedding $(1+3)$-dimensional spacetime.

As discussed with some of the authors of \cite{Tajik_2023},
experimentally there is the potential to measure the entanglement entropy 
in lab setups similar to network configurations discussed in this article, 
although technical limitations might restrict the complexity of the 
networks created. Alternative methods, like optical lattices, photonic integrated circuits or even materials like carbon nanotubes 
would offer more flexibility and allow for creating more complex networks. For example, quantum networks may also be realized in photon integrated circuits and hence give insights to their applications to optical quantum computers \cite{Andersen}, where, on a chip scale, squeezed states of light are feeded into an optical network consisting of several optical paths and beam splitters.

In the future, the increasing use of satellites with free space laser links 
for classical and quantum communication \cite{toyoshima2021recent, Liao:2017isj} opens up new possibilities 
for quantum network experiments and potentially gravitational wave 
detection similar to large scale classical experiments like LISA \cite{LISA:2017pwj}. Such networks build up by laser links as edges and satellites as vertices could be designed in various configurations, 
including fractal patterns inspired by fractal antennas in 
telecommunications with the advantage of having a high bandwidth and small size \cite{werner2003overview}. 

Throughout this article we presented graphs as an idealization of physical network structures on which quantum fields are confined. 
Extending this concept further, these networks might be thought as  
fundamental structures of nature itself. 
This perspective proposes a foundational role for $(1+1)$-dimensional physics,
suggesting that $(1+3)$-dimensional physics could be an 
emergent phenomenon on quantum networks. 
Envisioning quantum networks as intrinsic to the fabric of the Universe 
leads to a transformative approach to understanding physics. 
It implies that the complexities of our $(1+3)$-dimensional world 
might originate from simpler quantum processes within networks
embedded in a four-dimensional spacetime. $(1+3)$-dimensional physics would then be an effective theory with a UV cutoff given by the typical edge length $L$. Physics beyond this cutoff would then not be dictated by $(1+3)$-dimensional physics but by its $(1+1)$-dimensional counterpart and additionally influenced by the network topology.

Our approach, when compared with other methods, reveals distinct 
advantages when considering problems in QFT in curved spacetimes. 
For instance, it allows the use of conformal methods, 
unlike $(1+3)$-dimensional lattice QFT. 
These unique characteristics make our approach a 
valuable alternative in the study of quantum fields. 
The relationship between different vacuum states in $(1+3)$ dimensions 
and those on the network is another intriguing aspect. 
Investigations into phenomena like black hole formation and Hawking radiation near event horizons, 
where particles experience extreme blue shifts, 
suggest that local studies using quantum networks could offer valuable insights.
\label{concl}

\section*{Acknowledgments}
We want to express our gratitude to Erik Curiel, Ted Jacobson, John Preskill and Bill Unruh who took the time to discuss this idea with us and provided inspiring and valuable comments. It is a pleasure to thank Ivan Kukuljan and Mohammadamin Tajik for dedicating many hours to discuss in details our work, comparing with experimental results. 
Our thanks goes to Pablo Sala de Torres-Solanot for his thoughts and nice discussions on the topic and to Ka Hei Choi and Marc Schneider for fruitful comments and suggestions which helped with the presentation of the article.
The work of C.G. has been supported by the German Federal Ministry of Education and Research under a grant by the Begabtenförderungsnetzwerk.

\appendix*{}
	\section{}
\label{appendix}
\subsection*{Deconstructing the entanglement of continuous variable quantum systems}

Perhaps naively entanglement entropy is an ill-defined information measure 
in quantum field theory. The qualification refers to the implicit assumption 
of a classical spacetime which allows for coincidence limits.
In these limits entanglement entropy grows unbounded. 
Of course we do not know the structure of spacetime at arbitrary small scales. 
If entanglement entropy is a sensible concept in quantum field theory
then something must prevent coincidence limits. 
On the other hand, events are measured to happen 
at places as opposed to points, but experiments assign to each place a point 
by way of error estimation in accordance with a given resolution limit. 
The latter induces a discrete localization structure with a minimal distance
scale given by devices employed in the measurement. 
This resolution limit is a fact 
and cannot be removed in our case. 
Within the given framework, the entanglement entropy 
of quantum systems described by continuous variables can only 
be quantified relative to an extrinsically given discrete localization structure. 
The measurement device
effectively maps continuous variable quantum systems
to discrete quantum systems. We leave this investigation for future work. 

In this appendix we focus on the extreme distance hierarchy 
$a \ge 1/\mu=\xi$, where $a$ denotes the extrinsically induced minimal 
distance scale. Under the spell of this hierarchy field configurations 
confined to graphs are transformed to systems of finitely many
weakly coupled harmonic oscillators located on the edges of the graphs.
Clearly, hierarchies of this type entail the entire span from 
the weak coupling regime to the decoupling limit, that is, 
the perturbative domain allowing to consider the Hamiltonian 
of the system as a small deviation from its diagonal part. 
In this perturbative framework the entanglement entropy Eq.~\eqref{entrN}
can be calculated analytically \cite{Riera_2006, dimitrios}.

Consider again the single edge depicted in red in Fig.~\ref{meshGraph} 
supporting a field configuration subjected to Dirichlet 
boundary conditions at its endpoints.
Requiring that $a\gg 1/\mu$, instead of dealing with
a continuous variable quantum system, the system reduces 
to a one-dimensional chain of finitely many, say $N$, 
harmonic oscillators. We assume that an entangling sphere 
intersects this edge dividing into an interior region 
containing $n$ oscillators and an exterior region where
$N-n$ oscillators are located. In the interior and the exterior
region the oscillators are equidistant from each other. 
Recall the form of the matrix $K$ given in Eq.~\eqref{formK},
\begin{eqnarray}
    K
    =
    M^2 E
    + \Delta_\mathrm{l}
    \left(
        K
    \right)
    + \Delta_\mathrm{u}(K),
    \label{formK2}
\end{eqnarray}
where $E$ denotes the $(N\times N)$ unit matrix, 
$M^2 := \tfrac{1}{N}\mathrm{tr}(K)=2+\mu^2 a^2$,
and the lower- and upper-triangular matrix is
given by 
$(\Delta_\mathrm{l}(K))_{qs}
=
-\theta(N-s-1/2) \delta_{q,s+1}
$ and 
$
(\Delta_\mathrm{u}(K))_{qs}
=
-\theta(N-q-1/2) \delta_{q+1,s}$. 

In the formal extreme distance hierarchy with $\mu a \gg 1$
the diagonal term $\mathrm{tr}(K)$ dominates over the off-diagonal terms 
since $\Delta_\mathrm{l,u}(K)$ are $\mu a$-independent. 
This allows to expand $\Omega=\sqrt{K}$ around its diagonal contribution
\cite{Riera_2006} with off-diagonals suppressed by inverse powers
of $M$. The leading term in this expansion corresponds 
to the limit of decoupled oscillators characterized by a 
vanishing entanglement entropy. The onset of entanglement 
is encoded in the subleading terms,
\begin{eqnarray}
\Omega
=
M \, E - \tfrac{1}{2M} \, 
\left(\Delta_\mathrm{l}(K) + \Delta_\mathrm{u}(K)\right) 
+ \mathcal{O}\left(\tfrac{1}{M^3}\right)
\;. \, \, \,\, \,\, \,
\end{eqnarray}
In terms of the decomposition (\ref{decomOm}) of $\Omega$
into interior and exterior locations (relative to 
the entangling surface), we have 
$
(\Omega_\mathrm{II})_{qs} 
= 
M \delta_{q,s} 
- \tfrac{1}{2M} (\delta_{q+1,s} + \delta_{q,s+1})
+ \mathcal{O}(M^{-3})
$, with $q,s\in \lceil n \rfloor$.
The coefficients of $\Omega_\mathrm{EE}$ have the same 
form but with indices in $\lceil N-n \rfloor$.
Lastly, the submatrix $\Omega_\mathrm{IE}$ has a 
single nontrivial entry 
$-\tfrac{1}{2M} \delta_{u,n}\delta_{v,1}$
at leading order, with $u\in\lceil n\rfloor$, $v\in\lceil N-n\rfloor$, 
and $\Omega_\mathrm{EI}=(\Omega_\mathrm{IE})^\mathrm{T}$.

\begin{figure}[]
\centering
\includegraphics{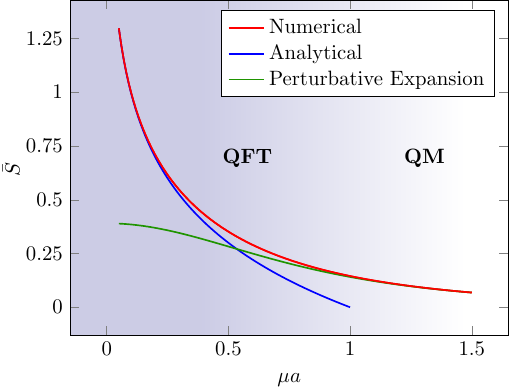}
	\caption{\label{rangeofmasses} 
    Entanglement entropy of degrees of freedom confined to a single edge 
    that pierces the entangling surface at its midpoint as a function of $\mu a$.
    The length of the edge allows for $500$ oscillator locations 
    equidistant from each other with separation $a$. This guarantees 
    $\mu a \gg 0.002$ in the displayed domain $\mu a \in (0.05, 1.5)$
    so that finite-size effects are not affecting the entanglement 
    entropy evaluated at the midpoint of the edge. 
    The red curve shows the numerical value obtained with 
    the code \cite{ourcode}, while the blue and green curve, 
    respectively, show $S^\mathrm{plat}$ and $S^{(3)}$
    given in the main text. The background shading indicates 
    the domain of validity of quantum field theory. }
\end{figure} 

Since the submatrix $\Omega_\mathrm{IE}$ encodes the correlations between 
degrees of freedom located in the interior and exterior 
relative to the entangling surface, it is a zero matrix 
in the decoupling limit and therefore the entanglement entropy 
vanishes. The leading deviation from the decoupling limit 
is given by the $\mathcal{O}(1/M)$ contribution in the $1/M$
expansion. This contribution introduces a nearest neighbor 
correlation and, in particular, a correlation across the entangling surface
between neighboring oscillators that are located in the interior and 
exterior, respectively, relative to this surface. 
Taking only the leading deviation from the decoupling limit 
in the $1/M$ expansion into account, the onset of entanglement 
across the entangling surface can be  
quantified by a straightforward computation of the entanglement 
entropy up to and including the $\mathcal{O}(1/M)$ terms 
of the expansion. 

If the entangling surface intersects an edge, the intersection
contains a single point. In the leading deviation from the 
decoupling limit, only one communication channel between the 
nearest neighbors located on both sides of the surface 
can be established to support correlations across the 
surface. 
Solving the eigenvalue problem for $\gamma^{-1}\beta $ up to leading
order \cite{dimitrios} in $1/M$ yields a single nonvanishing eigenvalue 
$\lambda= {1}/{8M^4}$. Inserting $\lambda$ 
into Eq.~\eqref{entrN} for the entanglement entropy gives,
\begin{equation}
    S^{(1)}= \frac{(1+4\ln(2M) )}{16M^4}\,,
\end{equation}
at leading order in the expansion. 
Including the next to and next-to-next-to-leading-order corrections
\cite{dimitrios}, 
\begin{equation}
\label{finalS1+1}
	{S}^{(3)}  =
    S^{(1)} + 
    \frac{1+328 \ln(2M)}{512M^8} 
	+ \frac{-599+5880 \ln(2M)}{3072M^{12}}\, .
\end{equation}
As could have been anticipated from our numerical experiments, 
sufficiently close to the decoupling limit the entanglement 
entropy does not depend on the radius of the entangling sphere. 
In particular, for the threshold value $M=\sqrt{3}$, Eq.~\eqref{finalS1+1} agrees quantitatively to $97\%$ with the numerical value shown in Fig.~\ref{loopgraph}. Moreover, for $M=\sqrt{3}$, 
entanglement across the entangling sphere is effectively restricted to
configurations involving up to three correlated nearest neighbors.
This agreement shows the validity (relative to the error) of 
the $1/M$ expansion even for the threshold value $M=\sqrt{3}$.

In Fig.~\ref{rangeofmasses} the entanglement entropy is depicted
as a function of $\mu a$ for a single edge piercing 
the entangling sphere so that the puncture in the intersection
coincides with the midpoint of the edge at $L/2$. 
The red curve represents the numerical value of the entanglement entropy. 
The blue curve shows the analytical result $S^\mathrm{plat}$ at $L/2$, 
as given in \cite{Calabrese_2004}.
Finally, the green curve shows $S^{(3)}$ as given 
in Eq.~\eqref{finalS1+1}.
As can be seen, the plateau function $S^\mathrm{plat}$
evaluated at $L/2$
is a good approximation for values of $\mu a$ up to $0.2$
provided the distance hierarchy $L\gg 1/\mu=\xi$ holds.
In Fig.~\ref{rangeofmasses} we chose $L=500 a$ which 
implies a lower bound $\mu a \gg 0.002$. 
For values $\mu a > 0.9$ the perturbative result $S^{(3)}$
is in good agreement with the numerical result.
Between these regimes, that is for $0.2 < \mu a <0.9$,
no analytical approximation is available (to our knowledge).

The $1/M$ expansion can be applied to any graph configuration.
For instance the \textit{loop} configuration considered 
in Sec.~\ref{subsecEE1p1} and depicted in blue in Fig.~\ref{meshGraph}.
The ($N\times N$)-matrix $K$ \eqref{formK} is adapted to accommodate periodic 
boundary conditions,
which relative to a finite resolution structure
are given by $\phi_0=\phi_N$ and $\phi_{N+1}=\phi_1$. 
In other words, $\phi_1$ and $\phi_N$ become nearest neighbors
and will establish an additional robust communication channel 
in the $1/M$ expansion. 
This is the only difference compared to the single-edge case, 
resulting in the following modification of the submatrix 
$\Omega_{\mathrm{IE}}$ of $\Omega=\sqrt{K}$:
$(\Omega_{\mathrm{IE}})_{uv}
=
-\tfrac{1}{2M}(\delta_{u,n}\delta_{v,1}+\delta_{u,1}\delta_{v,N-n})
$ to leading order in the $1/M$ expansion.
Compared to the single-edge case, the additional correlation
supported by the second term in $\Omega_{\mathrm{IE}}$
implies an extra eigenvalue $\lambda=1/8M^4$ different from zero, 
giving, as expected, an additional contribution to the 
entanglement entropy of the loop configuration. 
Since this contribution does not depend on $n$, the entanglement 
entropy of the loop configuration is twice the entanglement 
entropy of the single-edge (no loop) configuration, see
Fig.~\ref{loopgraph} for $\mu a=1$.

	\bibliography{lib.bib}
	
\end{document}